\documentclass[aps,prl,twocolumn,showpacs,10pt]{revtex4-1}
\usepackage{color}
\usepackage{epsfig}
\usepackage{multirow}
\usepackage{titlesec,bm}
\usepackage[dvipsnames]{xcolor}
\usepackage{graphicx,amsmath,amssymb,amsfonts,sidecap,color,hyperref}
\usepackage{subfigure}
\newcommand{\nn}{\nonumber\\}

\titleformat{\section}[runin]{\normalfont\itshape}{}{3pt}{}[.]
\begin{document}

\title{ Interaction Driven Floquet Engineering of Topological Superconductivity \\ in Rashba Nanowires}

\author{Manisha Thakurathi}
\author{Pavel P. Aseev}
\author{Daniel Loss}
\author{Jelena Klinovaja}

\affiliation{Department of Physics, University of Basel, Klingelbergstrasse 82, CH-4056 Basel, Switzerland}

\date{\today}

\begin{abstract}

We analyze, analytically and numerically, a periodically driven Rashba nanowire proximity coupled to an $s$-wave superconductor using bosonization and renormalization group analysis in the regime of strong electron-electron interactions.  Due to the repulsive  interactions, the superconducting gap is suppressed, whereas the Floquet Zeeman gap is enhanced, resulting in a higher effective value of $g$-factor compared to the non-interacting case. The flow equations for different coupling constants, velocities, and Luttinger-liquid parameters explicitly establish that even for small initial values of the Floquet Zeeman gap compared to the superconducting proximity gap, the interactions drive the system into the topological phase and the interband interaction term helps to achieve larger regions of the topological phase in parameter space. 
\end{abstract}

\maketitle

\section{ Introduction} 
In the past decade, a variety of experimentally relevant setups have been proposed to realize topological superconductivity~[\onlinecite{Alicea1}, \onlinecite{Beenakker}]. The exotic topological  phases include topological insulators~[\onlinecite{Kane,Zhang,Molenkamp,Leo,Amir,Hoti_Hsu}], Majorana fermions~[\onlinecite{Sarma,Oreg,Alicea2,Suhas1,Basel_exp,JK1,Simon,Vazifeh,Pientka,Ojanen,
Seo,Ruby,Mourik,Das,Deng,Churchill,Sen,Smitha}],  and parafermions~[\onlinecite{PF_Linder,PF_Clarke,PF_Cheng,PF_Mong,PF_Loss,PF_Oreg,PF_Vaezi,PF_Chen,PF_Klinovaja1,PF_Klinovaja2,PF_Fleckenstein,PF_Orth}].
Topological superconductors hosting Majorana modes provide a platform for topological qubits~[\onlinecite{Kitaev1}, \onlinecite{Kitaev2}], which have potential applications in quantum computation if missing quantum gates are supplemented~[\onlinecite{hoffman2016universal,plugge2016roadmap,karzig2017scalable}].  
Quantum systems driven out of equilibrium by an external field have also received rapidly increasing interest in recent years, for example, in the field of time-crystals~[\onlinecite{Wilczek,Nayak,Potter,Khemani}]. 
Combining these two research concepts gives rise to new phases such as Floquet Majorana modes~[\onlinecite{Fl_MT1,Fl_Kundu,Fl_MT2,Fl_JK,MT1,Fl_Paolo,Fl_Schoeller}],  Floquet topological insulators~[\onlinecite{Fl_Lindner,Fl_Katan,Fl_Kirill}], and higher order Floquet topological insulators~[\onlinecite{Fhoti_Liu,Fhoti_Sen,Fhoti_Yang}].

\begin{figure}[t]
\includegraphics[width=0.3\textwidth]{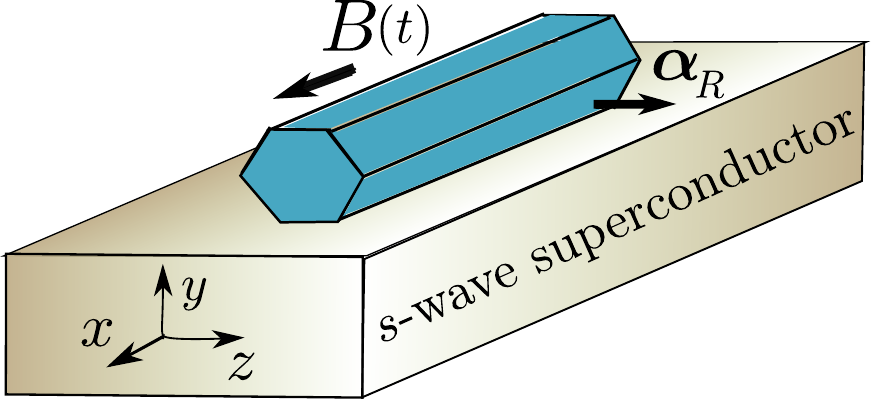}
 \caption{Sketch of the setup consisting of a one-dimensional Rashba nanowire (blue cylinder)  aligned along  $x$ direction and brought into proximity to an $s$-wave superconductor (yellow slab). The uniform time-dependent magnetic field $\textbf{\textit{B}}(t)$ with frequency $\omega$ is applied along the NW axis and, thus, is applied perpendicular to the SOI vector $\bm{\alpha}_R$.}
\label{fig_01}
\end{figure}

In this Letter, we present a comprehensive study of a setup consisting of an interacting Rashba nanowire (NW) proximity coupled to an $s$-wave superconductor and driven by a time-dependent magnetic field,
see Fig.~\ref{fig_01}.  This setup exhibits Floquet Majorana bound states (FMBSs) at each end of the NW if the Floquet Zeeman gap is larger than the superconducting  gap~[\onlinecite{MT1}].  However, high-amplitude time-dependent magnetic fields are not only difficult to apply but they also have detrimental effects on the superconductor, thus, the starting point of the current work is to focus on a low-amplitude magnetic field such that the Floquet Zeeman gap is small compared to the superconducting  gap. Moreover, in low-dimensional systems, electron-electron interactions are important as even for weak interaction strength, the system changes drastically and cannot be described as Fermi liquid anymore. A generic  interacting many-body system heats up to infinite temperature at sufficiently long times as described by the eigenstate thermalization hypothesis [\onlinecite{Rigol}]. However, as demonstrated in Ref.~[\onlinecite{Knap}], for a periodically-driven quantum many-body system, in a strong interaction regime, the prethermal Floquet state can be stabilized. Therefore, we consider prethermal region, which  means that the time period for the Floquet term is short compared to the heating time-scale and study the setup in the presence of electron-electron interaction~[\onlinecite{Suhas1}].  Using bosonization and renormalization group (RG) analysis~[\onlinecite{Giamarchi, Schoeller, MT2,Cardi,Senechal}], we show that even if the Floquet Zeeman gap is small compared to the superconducting gap, one can still obtain topological phases as the interaction renormalizes both superconducting and Floquet Zeeman terms. Repulsive electron-electron interactions renormalize  the Floquet Zeeman gap, which is of Peierls type [\onlinecite{Braunecker}], and make it larger, whereas  superconductivity gets suppressed such that the superconducting gap shrinks  [\onlinecite{Suhas1},\onlinecite{MT2}].
Thus, interactions helps to resolve the requirement of strong magnetic fields to obtain Floquet Majorana modes in the setup.

\section{Model}  \label{model1}
We consider a one-band Rashba nanowire (NW) aligned along $x$ direction, which is proximity coupled to an $s$-wave superconductor. The spin-orbit interaction (SOI) vector with strength $\alpha_R$ defines the quantization axis along $z$ direction. We apply an external time-dependent periodic magnetic field $\textbf{\textit{B}}(t)=B\,\cos(\omega\,t) \,\hat x$, with magnitude $B$ and frequency $\omega$, along the axis of the NW. This choice ensures that $\textbf{\textit{B}}(t)$ is perpendicular to the Rashba SOI vector. The Hamiltonian contains three terms, namely, the kinetic energy term and SOI term as $H_{kin}$, the superconducting pairing term $H_{sc}$, and the time-dependent Zeeman term $H_Z(t)$:
\begin{align}
&H(t)=H_{kin}+H_{sc}+H_Z(t),\\
&H_{kin}= \sum_\sigma \int dx\,\Psi_\sigma^\dagger(x) \left[\frac{ \hbar^2\hat k^2}{2\,m_0}-\mu \right] \Psi_\sigma(x)\nn&\hspace*{1.5cm}-\alpha_R \sum_{\sigma \sigma'} \int dx\,\Psi^\dagger_\sigma(x) (\sigma_z)_{\sigma \sigma'} \hat k\,\Psi_{\sigma'}(x),\nn
&H_{sc}=\frac{\Delta_{sc}}{2}\sum_{\sigma \sigma'}\int dx\, \Psi_{\sigma}(x)(i\,\sigma_y)_{\sigma \sigma'} \Psi_{\sigma'}(x),\nn
&H_Z(t)= 2\,t_F\, \cos(\omega\, t) \sum_{\sigma \sigma'} \int dx\,\Psi^\dagger_{\sigma}(x) \,(\sigma_x)_{\sigma \sigma'} \, \Psi_{\sigma'} (x), \nonumber
\end{align}
where $\Psi_\sigma(x)$ is the annihilation operator acting on an electron with spin $\sigma=\pm 1$ at position $x$, while the Pauli matrices $\sigma_{x,y,z}$ act on the spin space and $\hat k= -i \,\partial_x$ is the momentum operator. The amplitude of the time-dependent Floquet Zeeman term is given by $t_F=g\,\mu_B\,B/2$, where $g$ and $\mu_B$ are the $g$-factor and Bohr magneton, respectively. The proximity induced superconducting pairing gap is of the size $\Delta_{sc}$. The chemical potential $\mu$ is calculated from the SOI energy, $E_{so}=\hbar^2\,k_{so}^2/2m_0$, where $k_{so}=m_0\,\alpha_R/\hbar^2$ is the SOI wavevector and $m_0$ is the mass.

\begin{figure}[t]
\includegraphics[width=0.4\textwidth]{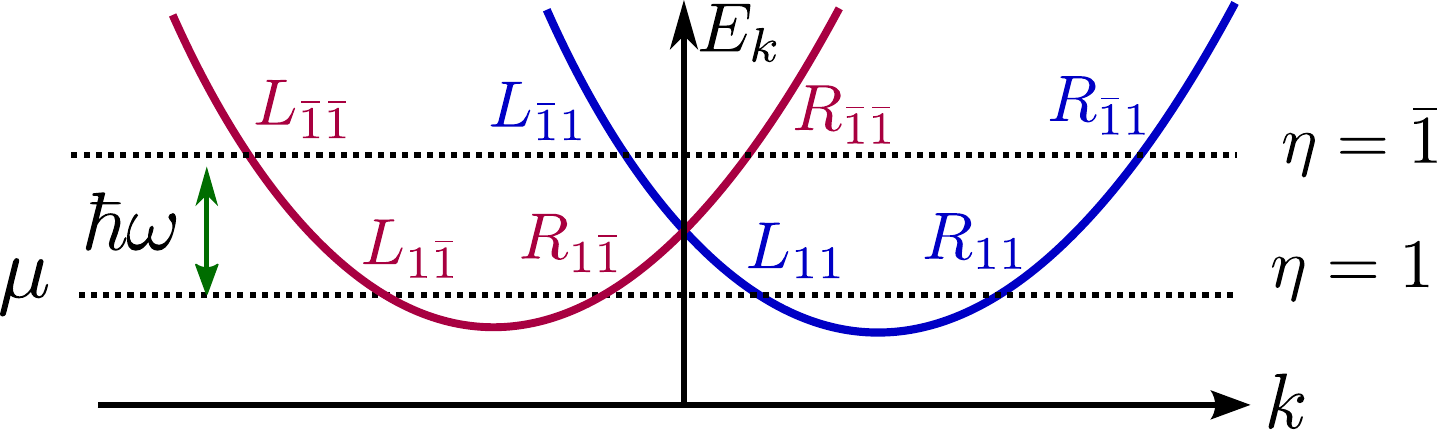}
\caption{Energy spectrum of Rashba NW consisting of  spin up ($\sigma=1$, blue) and spin down ($\sigma=\bar 1$, red) branches. The chemical potential $\mu$ is set below the SOI energy and the driving frequency $ \omega$  is chosen such that the smallest Fermi wavevector in the two Floquet bands, labeled by $\eta=\pm 1$, coincide. Close to the Fermi surfaces, the slowly varying right (left) fermionic field is denoted by $R_{\eta \sigma}$ ($L_{\eta \sigma}$).}
\label{spectra}
\end{figure}

We work in the Floquet formalism~[\onlinecite{Fl_MT1,Fl_Kundu,Fl_MT2, Fl_JK, Shirley}] to convert a time-dependent problem to a static one. The Hamiltonian is periodic in time, $H(t)=H(t+T)$, with period $T=2\,\pi/\omega$. The frequency $\omega$ is chosen such that the Floquet Zeeman terms become resonant.
As discussed in Ref. [\onlinecite{MT1}], this setup does not require the tuning of chemical potential to the SOI energy, instead the chemical potential $\mu$ has to be below the SOI energy such that the smallest Fermi wavevector in both effective bands coincide [see Fig. \ref{spectra}].  The eigenstates of Floquet operator,      
$H_F= H(t)-i\hbar\partial_t$, are given by periodic functions $e^{i n\omega t}$, where the integer $n$ labels different Floquet bands separated by an energy $\hbar\omega$. This periodicity allows us to write $H(t)=\sum_n H_n e^{in\omega t}$, where $H_n= \int_0^T  e^{-in\omega t} H(t)/T $. The Floquet Hamiltonian $H_F$  acquires a block diagonal form, written as
  \begin{align}
  H_F=  \begin{pmatrix}
\cdots &H_0+\hbar \omega & H_1 & 0 & \cdots\\
\cdots &H_{-1} & H_0 & H_{1} & \cdots  \\
\cdots &0 & H_{-1} & H_0 -\hbar \omega & \cdots
  \end{pmatrix},
  \end{align}
  where the static terms in $H(t)$, {\it i.e.} $H_0=H_{kin}+H_{sc}$,  act on each Floquet band, while the oscillatory magnetic field couples different Floquet bands which are separated by an energy $\hbar \omega$. We note that we work in the quasi-equilibrium limit such that we have a well-defined quasi-Fermi energy. 
 
 For simplicity, we consider only the single photon absorption processes, and thus focus only on the lowest two Floquet bands denoted by $\eta=1$ and $\bar 1$. Thus, introducing the electronic operator $\Psi_{\eta \sigma}$, which annihilates an electron of spin $\sigma$ in band $\eta$, the Floquet Hamiltonian in the basis $\chi_{\eta\sigma}$=($\Psi_{11}$, $\Psi_{1\bar 1}$, $\Psi^\dagger_{11}$, $\Psi^\dagger_{1\bar 1}$, $\Psi_{\bar 1 1}$, $\Psi_{\bar 1\bar 1}$, $\Psi^\dagger_{\bar 11}$, $\Psi^\dagger_{\bar1\bar 1}$) is written as 
  \begin{align}
  H'_F=&\int dx\, \chi^\dagger_{\eta \sigma} \Big[\left(\frac{\hbar^2\hat k^2}{2m_0}-\mu -\alpha_R\sigma_z \hat k \right)\tau_z + \Delta_{sc}\tau_y \sigma_y\nn
  &+ t_F \tau_z \sigma_x \eta_x +\frac{\hbar\omega}{2}{(1-\eta_z)\tau_z}\Big]\chi_{\eta\sigma} .
  \label{HFlo}
  \end{align}
Here the Pauli matrices $\tau_{x,y,z}$ and $\eta_{x,y,z}$ act on particle-hole and Floquet band spaces, respectively.
If the Floquet Zeeman gap exceeds the superconducting pairing gap, $t_F>\Delta_{sc}>0$, the system hosts two zero-energy bound states protected by an effective time-reversal symmetry $\mathcal{T}=-i \eta_z \sigma_y \mathcal{K}$, where $\mathcal{K}$ is the complex conjugation operator [\onlinecite{MT1}].

Next, we include electron-electron  interaction.
First, we rewrite the kinetic energy in terms of the bosonic fields [see Supplemental Material (SM) [\onlinecite{SM}]] as
\begin{align}
&H_{kin}'= \sum_{\eta \beta}  \int \frac{dx}{2\pi} \Bigg[
u_{\eta \beta} \Big[\frac{\left ( \partial_x \phi_{\eta \beta}  \right )^2 } {K_{\eta \beta}} + K_{\eta \beta} \left (\partial_x \theta_{\eta \beta}  \right )^2\Big] \nonumber  \\ 
&\hspace*{3.5cm}+2\,\tilde u\left( \partial_x \phi_{1c} \right) \left( \partial_x \phi_{\bar{1}c} \right)\Bigg],\label{bHkin}
\end{align} 
where the indices of the bosonic fields $\phi_{\eta\beta}$ and $\theta_{\eta\beta}$ refer to $\beta=c\, (s)$ for charge (spin) sectors of the $\eta$-Floquet band. We set $\hbar=1$  here as well as in the following calculation. Further, $u_{\eta \beta}$ and $K_{\eta \beta}$ corresponds to the velocity and Luttinger Liquid (LL) parameter, respectively. The cross-term, characterized by the velocity $\tilde u$,  describes the repulsive interaction between the Floquet bands. To simplify the problem, we fix $K_{1\beta}=K_{\bar 1 \beta}=K_{\beta}$ and $u_{1 \beta}=u_{\bar 1\beta}=u_{\beta}$. For an ideal LL,  $u_{\beta}/v_{F}=1/K_{ \beta}$ and $\tilde u/v_{F}= 1/K_{ c}^2 -1$, where $v_F$ is the Fermi velocity [\onlinecite{DL2}, \onlinecite{DL1}].  Notably, we consider only the charge density interaction between the Floquet bands, as we are interested in the parameter regime in which interactions are spin-rotational symmetric with $K_s \sim 1$. The superconducting pairing term and the Floquet Zeeman term are rewritten as
\begin{align}
H'_{sc}&=\sum_\eta \int dx\,\frac{2\,\Delta_{sc}}{\pi \, \alpha} \cos (\sqrt{2}\,\theta_{\eta c}) \cos(\sqrt{2}\,\phi_{\eta s}),\label{bHsc}\\
H'_Z&=\sum_\eta \int dx\,\frac{t_F}{\pi \, \alpha} \cos(\tilde \varphi_\eta/\sqrt{2}),
\label{bHZ}
\end{align}
where  
$\tilde \varphi_\eta=\phi_{\eta c}-\theta_{\eta c}-\phi_{\eta s}+\theta_{\eta s}+\phi_{\bar \eta c}+\theta_{\bar \eta c}+\phi_{\bar\eta s}+\theta_{\bar\eta s}$ and $\alpha$ is the renormalized lattice constant of the NW, which grows under RG.
As a result, the total effective Hamiltonian is given by $H'= H'_{kin}+ H'_{sc}+H'_Z$. 

\section{RG equations and analysis} Next, we derive the RG equations for different  coupling constants, velocities, and LL parameters in $H'$.    Most importantly, we are interested in finding out whether it is possible to reach the topological phase  even if the initial (non-renormalized) value of $t_F$ is smaller than $\Delta_{sc}$ and, thus, one expects the system to be in the trivial phase in the absence of interactions. To compare different competing terms, we work in  dimensionless units and define $\tilde \Delta_{sc}= \alpha\, \Delta_{sc}/u_c$ and $\tilde t_F=\alpha\,t_F/u_c$. Further, performing an RG analysis, our goal is to determine when $\tilde t_F$ dominates over $\tilde \Delta_{sc}$. In RG language, this means finding the parameter regime when  the system reaches the strong coupling limit, {\it i.e.} $\tilde t_F \sim 1$ and $\tilde t_F>\tilde \Delta_{sc}$. 

\begin{figure}[t]
\begin{center} \begin{tabular}{cc}
\epsfig{figure=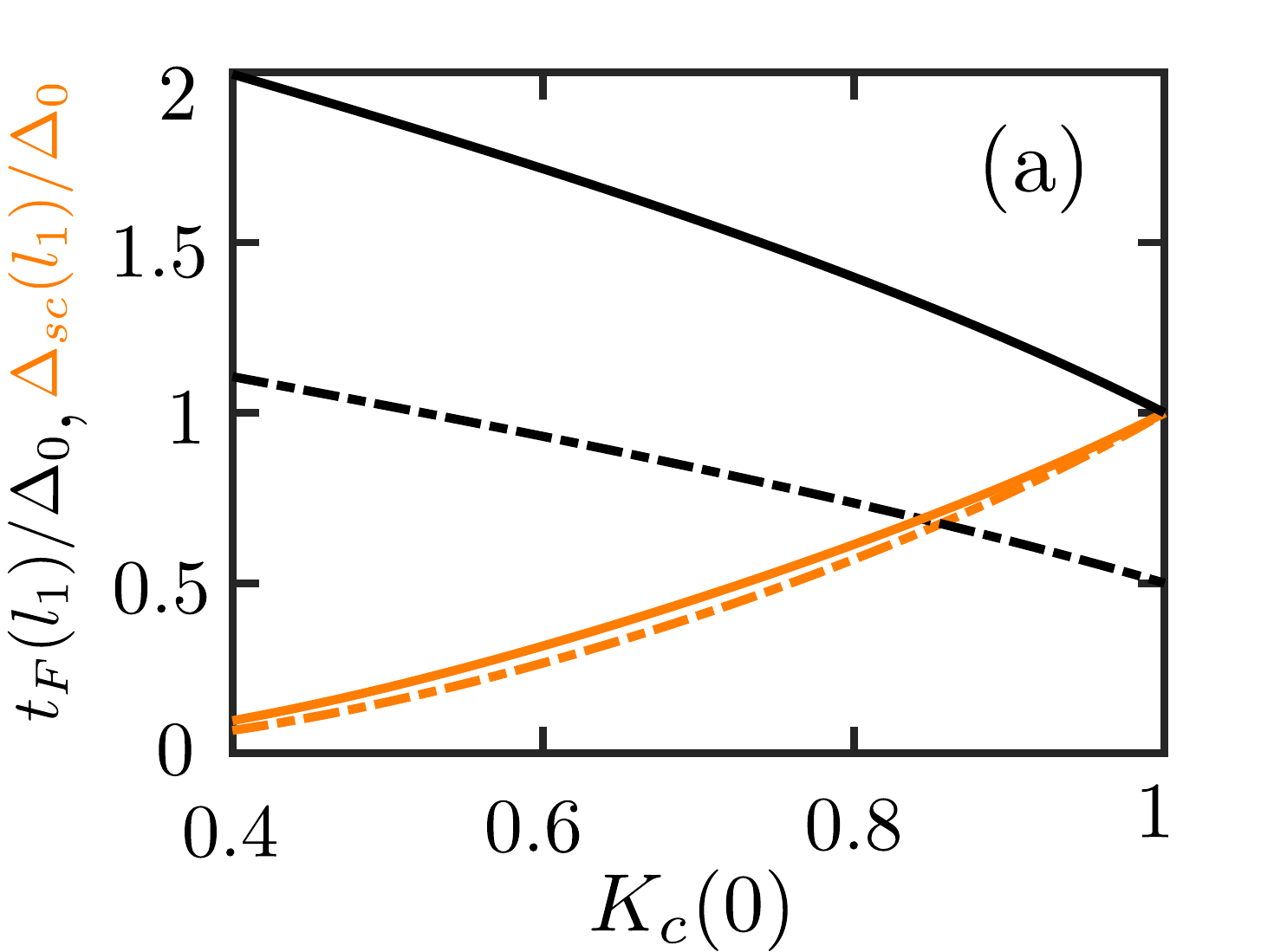,width=1.78in,height=1.44in,clip=true} &\hspace*{-0.5cm}
\raisebox{-0.25mm}
{\epsfig{figure=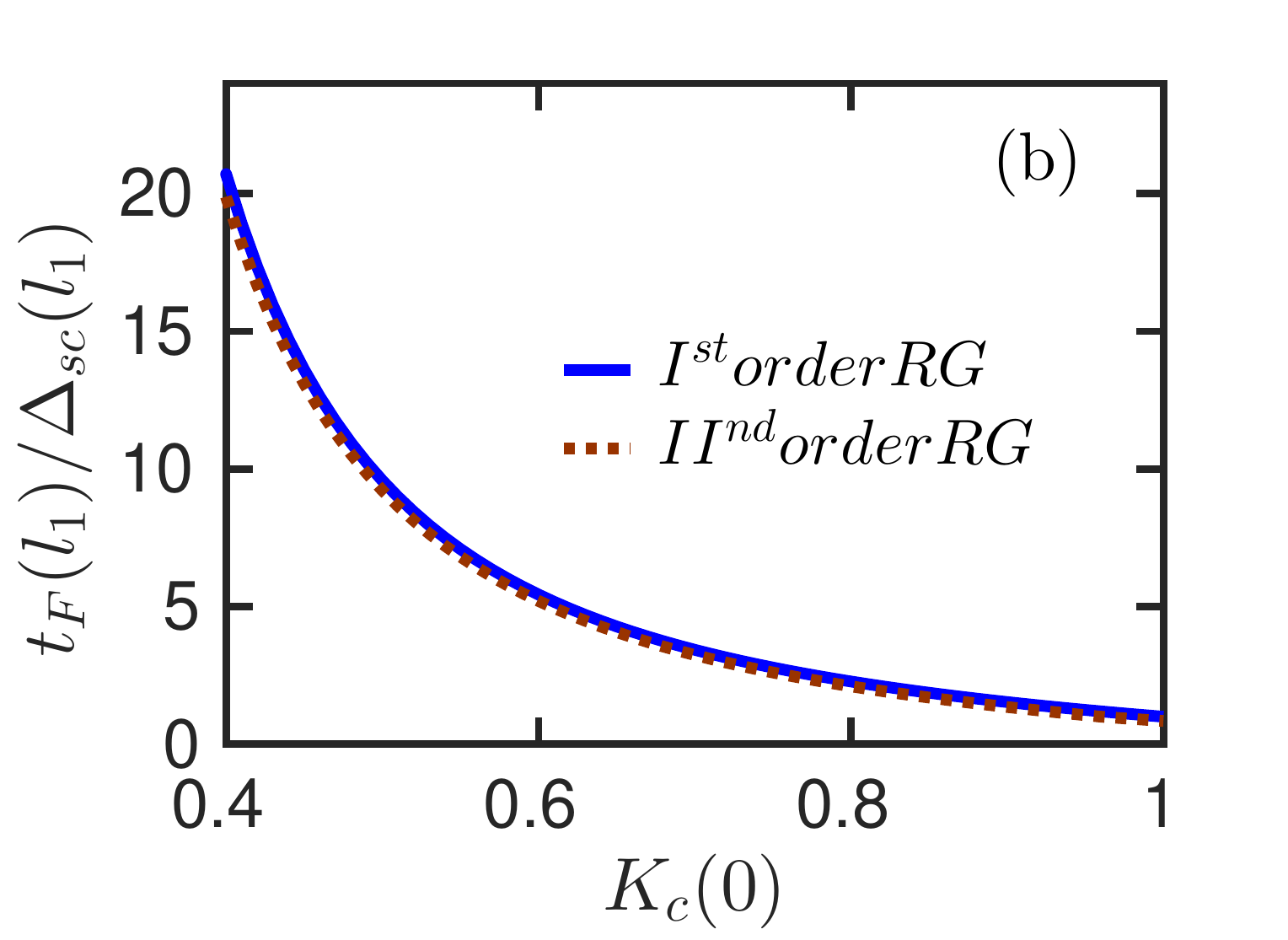,width=1.78in,height=1.47in,clip=true}}
\end{tabular} \end{center}
\caption{(a) The strength of the physical Floquet Zeeman term $t_F(l_1)/\Delta_0$ (black) and superconducting pairing term $\Delta_{sc}(l_1)/\Delta_0$ (orange) as a function of $K_c(0)$  for two initial values of  $t_F(0)/\Delta_{sc}(0)=0.5$ (dashed) and 1 (solid) for $\Delta_0=0.01\, u_c(0)/a_0$ [see Eq. (\ref{ana})]. The system is in the topological phase when $t_F(l_1)>\Delta_{sc}(l_1)$. As the ratio $t_F(0)/\Delta_{sc}(0)$ decreases, we need stronger interactions to reach the topological phase. (b) To compare the results of first-order RG [blue curve, see Eq. (\ref{ana})] and second-order RG (red curve, see SM [\onlinecite{SM}]) flows, we plot the ratio $t_F(l_1)/\Delta_{sc}(l_1)$ as  function of initial value of $K_c$ for $t_F(0)/\Delta_{sc}(0)=1$.  The second-order RG gives only small corrections to the first-order RG results. The correction is  most relevant close to the point $K_c=1$, where analytical results predict that the system is topological [$t_F(l_1)>\Delta_{sc}(l_1)$], however,  due to the second-order correction, the system is in the trivial phase. Other parameters used in the second-order RG numerics are $K_s(0)=1$, $u_c(0)/v_F=1/K_c(0)$, $u_s(0)/v_F=1$, and $\tilde u(0)/v_F=1/K_c^2(0)-1$.}
\label{Fig_04}
\end{figure}

To derive the RG equations, first we calculate different correlation functions between $\phi_{\eta \beta}$ and $\theta_{\eta \beta}$ for the kinetic part  $H'_{kin}$ using Green functions [\onlinecite{SM}].
Subsequently, utilizing the operator product expansion (OPE) method [\onlinecite{Cardi}], we compute the RG equations up to the first-order, 
\begin{align}
&d\tilde \Delta_{sc}/dl= \;\left[2-(\lambda_+ +\lambda_-+ K_s)/2\right]\,\tilde \Delta_{sc},\nn 
&d\tilde t_F/dl= \;\left[2-( 2\,\nu_++ 2\,\lambda_- + \nu_s )/4\right]\, \tilde t_F. 
\label{RGE}
\end{align}
Here, we define 
$\lambda_{\pm}= (u_c \pm \tilde{u}\,K_c)/(2\,K_c\,u_\pm)$, $u_{\pm}^2 =\,u_c^2 \pm u_c \, \tilde{u}\, K_c$, \,$\nu_s= K_s+ 1/K_s$, and
$\nu_+= K_c\, u_c/2\,u_+$. The dimensionless RG flow parameter is $l= \text{ln}(\alpha/\alpha_0)$, where $\alpha_0$ is the bare value of the lattice constant. Notably, the velocities and LL parameters do not flow under first-order RG. We are interested in the gapped regime where  $\tilde \Delta_{sc}$  and  $\tilde t_F$ are RG relevant (terms growing as a function of $l$). Moreover, the superconducting pairing term $\tilde \Delta_{sc}$ ($\tilde t_F$) is RG relevant for $\lambda_+ + \lambda_- +K_s<4$ ($2\nu_+ + 2\lambda_- +\nu_s<8$). To estimate the relevant parameter regime for $K_c$,  we consider the limiting case $\tilde u=0$ and $K_s=1$, which results in  $K_c>1/3$ ($K_c +1/K_c <6$) for $\tilde\Delta_{sc}$ ($\tilde t_F$) to be RG relevant, thus providing us the lower bound. Therefore, in what follows, we focus on the repulsive interaction regime with $1/3<K_c<1$. 

We solve the first-order RG [see Eq. (\ref{RGE})] for $\tilde \Delta_{sc}$ and $\tilde t_F$ considering that the latter reaches the strong coupling limit {\it i.e.} $t_F(l_1)=1$, where $l_1$ is the dimensionless RG flow parameter. 
Therefore, one obtains $l_1= 4\, \text{ln} \,\tilde t_F (0)/(2\,\nu_++ 2\, \lambda_-+ \nu_s -8)$ and
$\tilde \Delta_{sc} (l_1)=\tilde \Delta_{sc}(0)\, [\tilde t_F(0)]^\gamma$, where $\gamma=(4- \lambda_+-\lambda_--K_s)/(\nu_++  \lambda_-+ \nu_s/2 -4)$.
For $K_s=1$, the values of the physical superconducting pairing gap and Floquet Zeeman gap are given by
\begin{align}
&\Delta_{sc}(l_1)= \,\frac{u_c(0) \, \tilde \Delta_{sc}(0)}{\alpha_0} \,\left[\tilde t_F(0)\right]^{\frac{\lambda_++ \lambda_--1}{3-\nu_+- \lambda_-}}, \nn
& t_F(l_1)=  \,\frac{u_c(0)}{\alpha_0} \,\left[\tilde t_F(0)\right]^{\frac{2}{3-\nu_+- \lambda_-}},\nn
 &\nu_++\lambda_-= \left(K_c(0)/\sqrt{2-K_c^2(0)}+1\right)/2,\nn
&\lambda_++\lambda_-=  \left( \sqrt{2-K_c^2(0)}/K_c(0)+1\right)/2.
\label{ana}
\end{align}

In the presence of strong repulsive interaction, the Floquet Zeeman gap tends to exceed the superconducting pairing gap even if this was not the case for the initial values, see Fig. \ref{Fig_04}(a). Thus, interaction drives the system into the topological phase by satisfying the criterion $t_F>\Delta_{sc}$.  Generally, there is a crossover between $t_F$ and $\Delta_{sc}$ depending upon the ratio of their initial values.
When $t_F(0)/\Delta_{sc}(0)=1$, the crossover  happens at $K_c=1$ and thus for $K_c<1$ ($K_c>1$), the system is in topological (trivial) phase.  As the ratio $t_F(0)/\Delta_{sc}(0)$ decreases, the crossover point for $K_c$, which can be calculated by putting $\Delta_{sc}(l_1)=t_F(l_1)$ in Eq. (\ref{ana}),  shifts to  smaller values,  indicating that one requires stronger repulsive interaction in the NW to reach the topological phase.

\begin{figure}[t]
 \begin{tabular}{cc}
\epsfig{figure=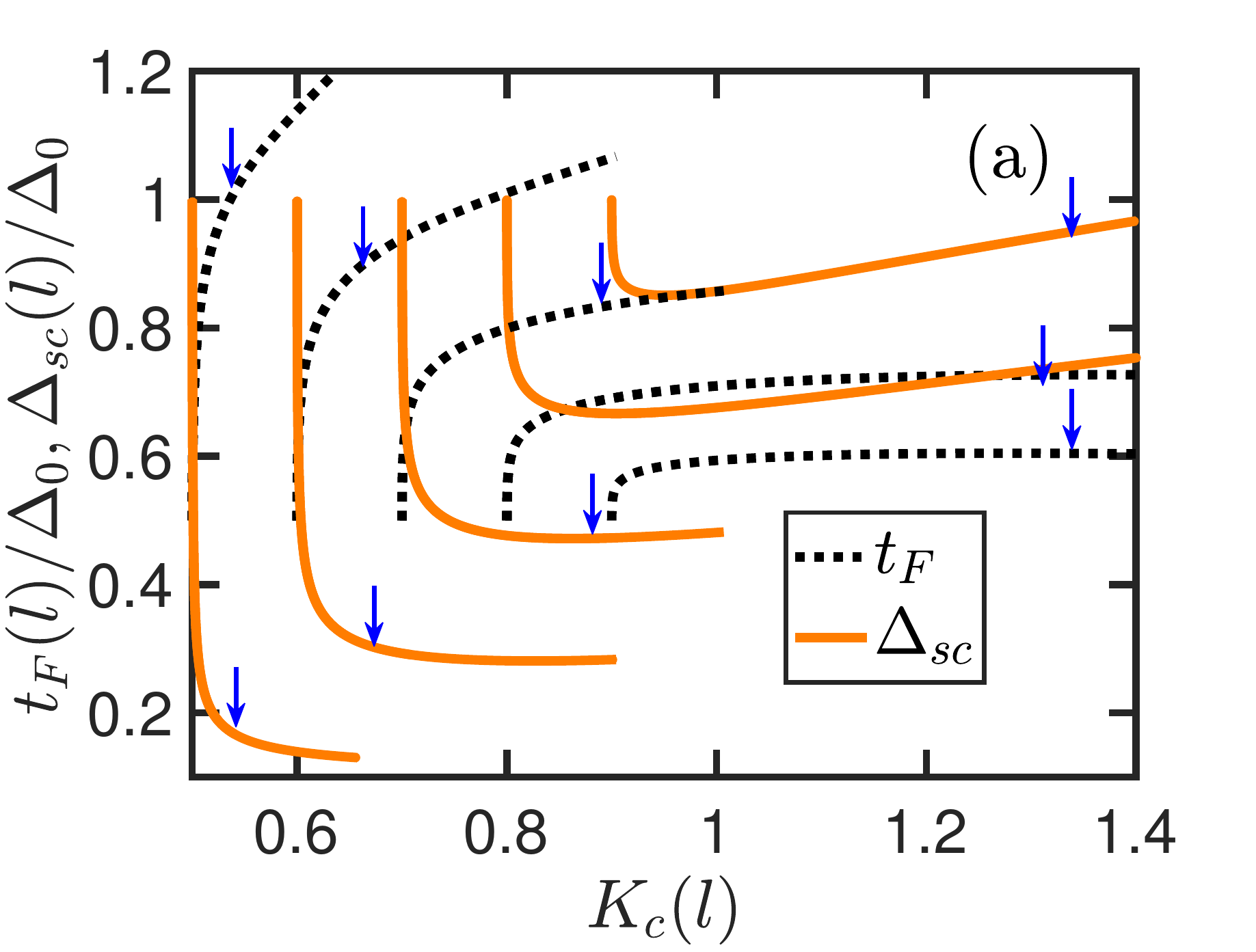,width=1.8in,height=1.47in,clip=true} &\hspace*{-0.5cm}
\raisebox{-0.25mm}
{\epsfig{figure=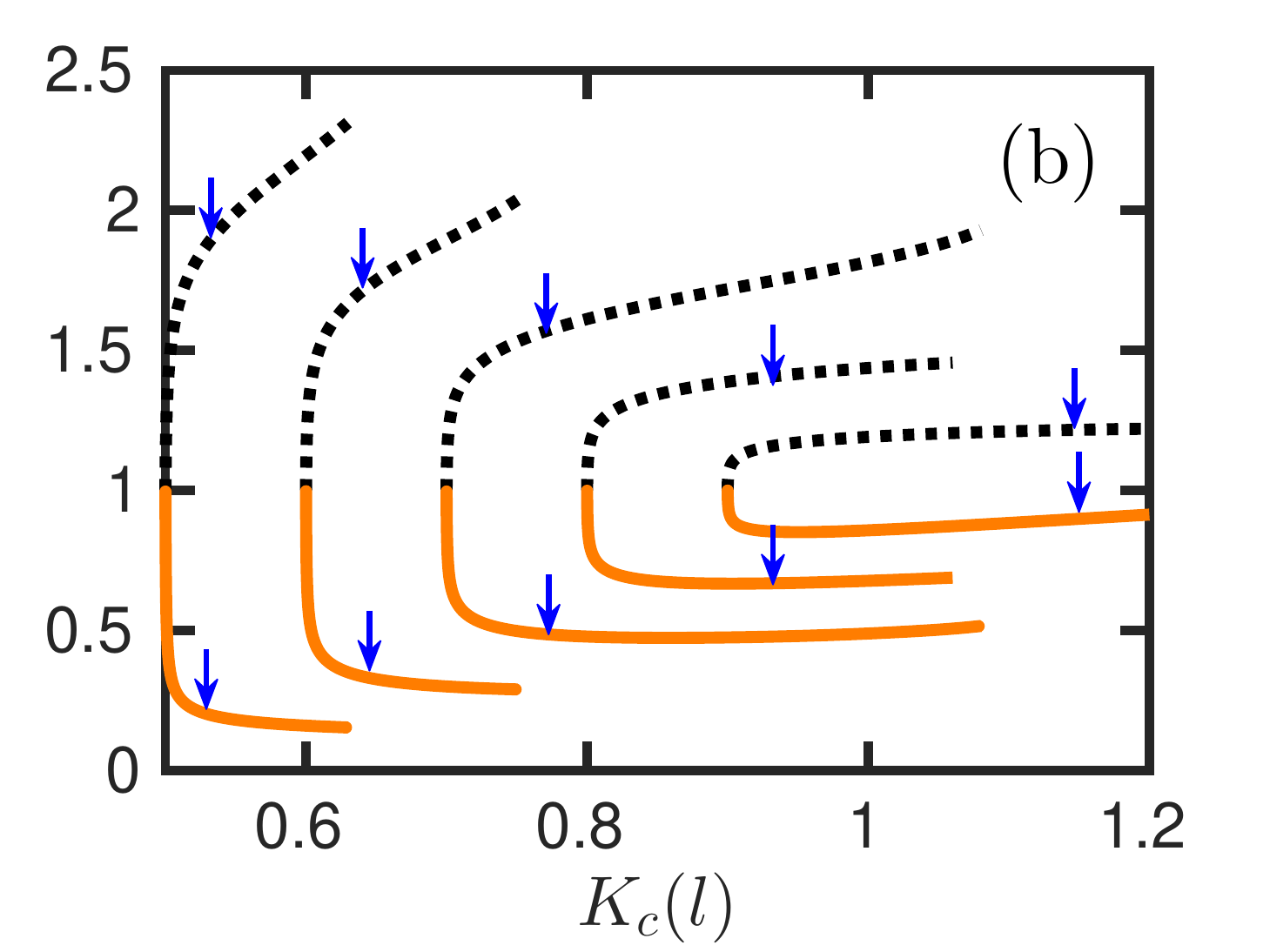,width=1.78in,height=1.47in,clip=true}}
\end{tabular} 
\caption{RG flow of the physical gaps $\Delta_{sc}/\Delta_0$ (orange solid) and $t_F/\Delta_0$ (black dotted) as a function of LL parameter $K_c$ after solving the second-order RG equations for (a) $t_F(0)/\Delta_{sc}(0)=0.5$ and (b)  1. The vertical blue arrow denotes the point when $\tilde t_F=1$. For the values $K_c(0)>0.8$ [$K_c(0)>1$], the system is  in the trivial phase for (a) [(b)]. The LL parameters stay close to the initial value before reaching the blue vertical arrow, which justifies the assumption of first order RG. The smaller the ratio $t_F(0)/\Delta_{sc}(0)<1$, the stronger interactions are required to reach the topological phase. Other initial conditions are chosen as $K_c(0)=0.5, 0.6, 0.7, 0.8, 0.9$ from left to right, $K_s(0)=1$, $u_c(0)/v_F=1/K_c(0)$, $u_s(0)/v_F=1$, and $\tilde u(0)/v_F=1/K_c^2(0)-1$. }
\label{fig_07}
\end{figure}

To check that the flow of LL parameters and velocities do not affect the analytical results obtained in the first-order, we recompute the RG equations up to the second-order [\onlinecite{SM}]. 
As these are involved coupled differential equations,  we solve them numerically. Generally, the second-order RG eqs. give small corrections to the first-order result, see Fig. \ref{Fig_04}(b). However, close to $K_c=1$, the corrections are more relevant rendering the system  trivial. Examining the RG flow of the physical gaps in second-order [see Fig. (\ref{fig_07})], we observe that, up to the strong coupling limit point, $K_c$ stays close to the initial value and hardly flows. This justifies our assumption of focusing only on the first-order RG equations. Also the effect of the ratio $t_F(0)/\Delta_{sc}(0)$ on the physical gaps in both RG orders matches exactly [see Figs. \ref{Fig_04}(a) and \ref{fig_07})]. Generally, the repulsive interaction suppresses the superconducting gap [\onlinecite{Suhas1},\onlinecite{MT2}]. In contrast to that, the Floquet Zeeman gap is enhanced [\onlinecite{Braunecker}], which results in a higher value of the effective $g$-factor compared to the non-interacting case. One can understand the enhancement in a simple way, the Floquet Zeeeman term 
has a form similar to a spin-flip backscattering term. As discussed in Ref. [\onlinecite{Giamarchi}], the backscattering amplitude increases as interactions get stronger, thus resulting  larger gaps in comparison to the non-interacting case. This allows one to satisfy the  topological criterion $t_F(l_1)>\Delta_{sc}(l_1)$ even if the non-interacting bare value of $t_F(l_0)$ is smaller than $\Delta_{sc}(l_0)$. 
\begin{figure}[b]
\begin{center} \begin{tabular}{cc}
\epsfig{figure=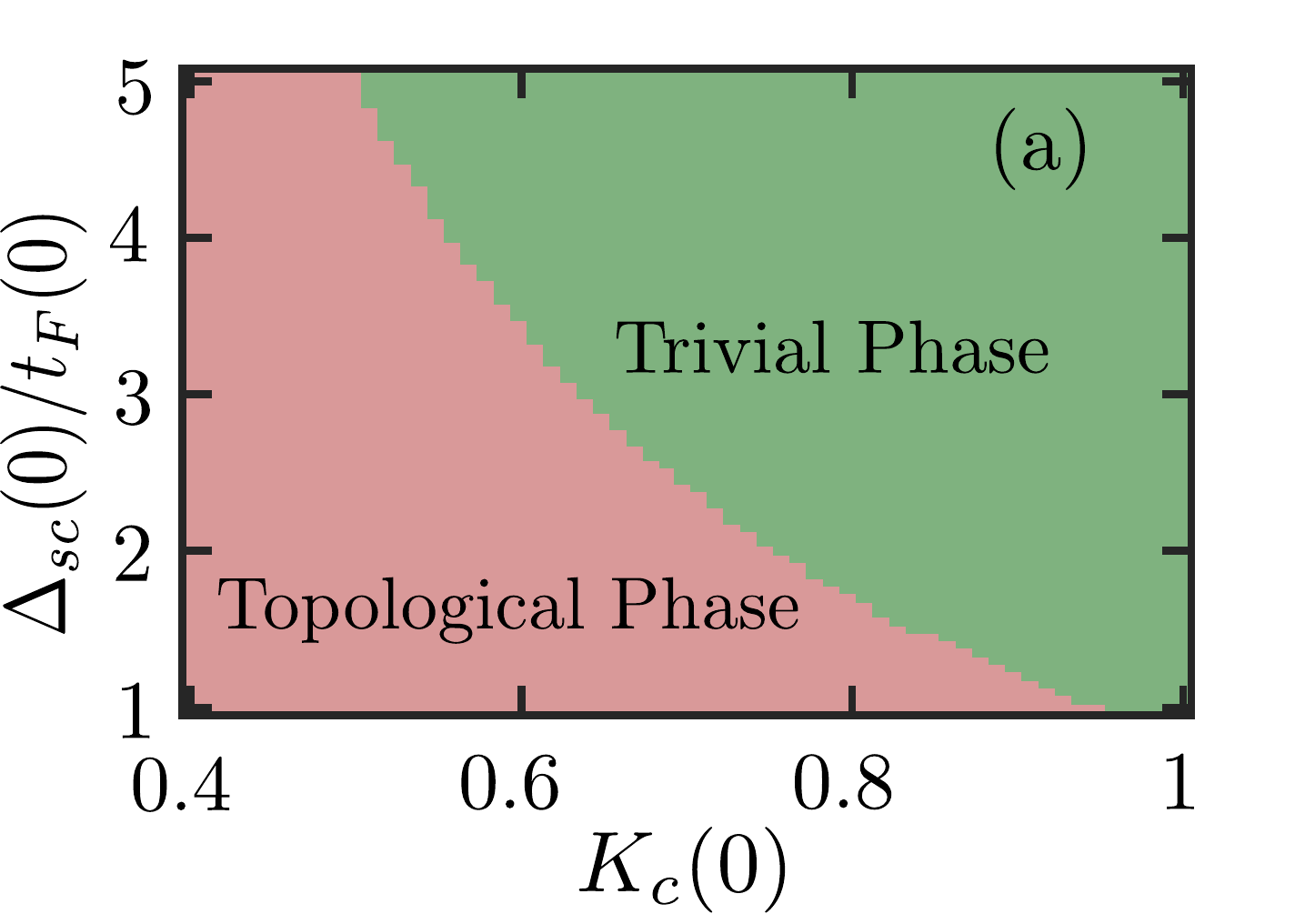,width=1.78in,height=1.43in,clip=true} &\hspace*{-0.5cm}
\raisebox{0.5mm}{\epsfig{figure=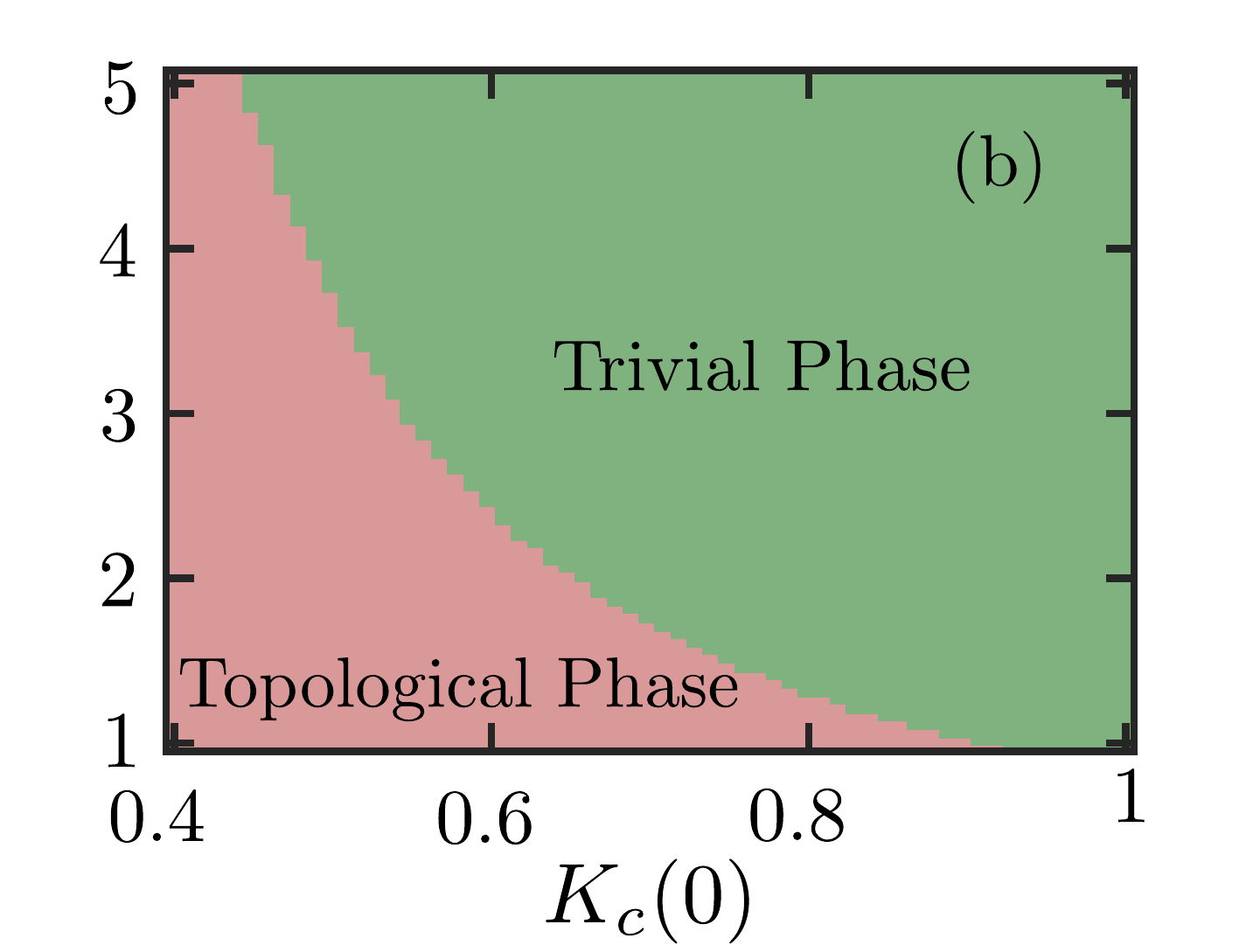,width=1.78in,height=1.41in,clip=true} }
\end{tabular} \end{center}
\caption{(a) Phase diagram as a function of $\Delta_{sc}(0)/t_F(0)$ and $K_c(0)$  in the presence  of the interband interaction cross-term, $\tilde u(0)/v_F=1/K_c^2(0)-1$, obtained numerically by solving second-order RG equation (see SM \onlinecite{SM}). (b) The same in the absence of $\tilde u(0)$.
If the topological criterion $t_F> \Delta_{sc}$ is satisfied, the system is in the topological phase (red area), otherwise in the trivial phase (green area). The area corresponding to the topological phase is substantially increased in the presence of $\tilde u$. Other initial conditions: $  K_s(0)=1$, $u_c(0)/v_F=1/K_c(0)$, and $u_s(0)/v_F=1$.}
\label{Phase}
\end{figure}

Finally, we also compute the phase diagram as a function of the ratio  $\Delta_{sc}(0)/t_F(0)$  and  the initial  LL parameter value $K_c(0)$ [see Fig. \ref{Phase}]. For a non-interacting system ($K_c=1$), if $\Delta_{sc}(0)/t_F(0) \ge 1$, system is always in the trivial phase. However, when $K_c<1$, the topological phase emerges due to the presence of interactions. As the ratio $\Delta_{sc}(0)/t_F(0)$ increases, we require lower values of $K_c$, {\it i.e.} stronger interaction strength to reach the topological phase. We also would like to emphasize the role played by the interband interaction cross-term  $\tilde u$. If $\tilde u$ is absent [see Fig. \ref{Phase} (b)], the phase boundary between the topological and trivial phase is shifted to lower values of $K_c(0)$ compared to the case of finite $\tilde u$ [Fig. [\ref{Phase} (a)]. Thus, the interband interaction term results in larger parameter space corresponding to the topological phase.

\section{Conclusions} We studied the effects of electron-electron interactions on a driven Rashba NW with proximity gap and analyzed the interplay between the Floquet Zeeman term and superconducting pairing term using bosonization techniques and RG analysis. The repulsive Coloumb interaction drives the system into the topological phase even if the initial (bare) value of the Floquet Zeeman gap is smaller than the superconducting proximity gap. Under RG flow, the physical Floquet Zeeman gap is enhanced whereas the proximity gap is suppressed, pushing the system into  the topological phase. The proposed setup is important as it does not require the tuning of the chemical potential close to the spin-orbit energy and it exhibits topological superconductivity, and thus Floquet Majorana modes, even for weak strengths of the driving magnetic field due to the presence of electron-electron interactions.
\label{con}

{\it Acknowledgments} --This work was supported by the Swiss National Science Foundation (SNSF) and NCCR QSIT. This project received funding from the European Union’s Horizon 2020 research and innovation program (ERC Starting Grant, grant agreement  No 757725).

\onecolumngrid
\newpage
\vspace*{1cm}
\begin{center}
    \large{\bf Supplemental Material:  Interaction driven Floquet engineering of Majorana modes \\}
\end{center}
\begin{center}
   Manisha Thakurathi, Pavel P. Aseev,  Daniel Loss, and Jelena Klinovaja\\
    {\it Department of Physics, University of Basel, Klingelbergstrasse 82, CH-4056 Basel, Switzerland}
\end{center}

\setcounter{section}{0}
\setcounter{equation}{0}
\setcounter{figure}{0}
\setcounter{page}{1}
\makeatletter
\renewcommand{\thesection}{S\arabic{section}}
\renewcommand{\theequation}{S\arabic{equation}}
\renewcommand{\thefigure}{S\arabic{figure}}
\titleformat{\section}[hang]{\large\bfseries}{\thesection.}{5pt}{}

 \section{Bosonization}
\label{bs}
 In this section, we first linearize the spectrum close to the Fermi momenta and subsequently bosonize the Hamiltonian in order to include electron-electron interactions in the analysis [\onlinecite{Giamarchi, Schoeller, MT2,Cardi,Senechal,Shirley}].  The Fermi points $k_{F, \eta\sigma r}$ of the Floquet band $\eta$ with the spin $\sigma$  have the form $k_{F,{1\sigma \pm}}=\sigma\,k_{so}\pm k_{so}\sqrt{1+\mu/E_{so}}$ and $k_{F,{\bar 1\sigma \pm}}=\sigma\,k_{so}\pm k_{so}\sqrt{1+(\mu+\hbar \omega)/E_{so}}$. The resonance condition is satisfied, if $k_{F,1\bar 1 +}=k_{F,{\bar 1 1 -}}$ and $k_{F,{\bar 1 \bar 1 +}}=k_{F,{11 -}}$. We write the fermionic fields in terms of the right mover and left mover fields as
 \begin{align}
 \Psi_{11}&= R_{11}e^{i k_{F,{11+}}}+ L_{11}e^{i k_{F,{11-}}},\nn
 \Psi_{1\bar 1}&= R_{1\bar 1}e^{i k_{F,{1\bar 1+}}}+ L_{1\bar 1}e^{i k_{F,{1\bar 1-}}},\nn
 \Psi_{\bar 1 1}&=R_{\bar 1 1}e^{i k_{F,{\bar 11+}}}+ L_{\bar 11}e^{i k_{F,{\bar 11-}}}, \nn
 \Psi_{\bar 1 \bar 1}&= R_{\bar 1 \bar 1}e^{i k_{F,{\bar 1\bar 1+}}}+ L_{\bar 1\bar 1}e^{i k_{F,{\bar 1\bar 1-}}}.
 \end{align}
 
 Here we denote the slowly-varying right and left moving field by $R(x)$ and $L(x)$. Further, we linearize the sum of kinetic energy and SOI terms, which takes the following form 
 \begin{align}
\hspace*{-0.4cm} H'_{kin}=-i\,\hbar \, v_{F}\sum_{\eta \sigma} \int dx\,[R_{\eta \sigma}^\dagger(x) \partial_x R_{\eta \sigma}(x)- L_{\eta \sigma}^\dagger(x) \partial_x L_{\eta \sigma}(x)],
 \end{align}
where $v_F$ is the Fermi velocity in the NW. The linearized form of $s$-wave pairing term has the following form 
\begin{align}
H'_{sc}=\Delta_{sc}\sum_\eta \int dx\,[R_{\eta \bar 1}^\dagger L_{\eta 1}^\dagger - R_{\eta 1}^\dagger L_{\eta \bar 1}^\dagger + \text{H.c.}].
\end{align}
Finally, in terms of fermionic right and left movers, the Floquet Zeeman term that couples the lower and upper  Floquet bands is given by
\begin{align}
H'_Z= t_F \sum_\eta \int dx\,[R_{\eta \bar 1}^\dagger L_{\bar \eta 1} + \text{H.c.}].
\label{spinFlip}
\end{align}

To include the electron-electron interactions, we consider only the low-lying excitations close to the Fermi level. As the particle-hole excitation are bosonic in nature, we bosonize the Hamiltonian by defining left and right moving fermions in terms of the charge ($\phi_{\eta c},\theta_{\eta c}$) and spin ($\phi_{\eta s},\theta_{\eta s}$) bosonic fields described by following definition 
\begin{align}
R_{\eta \sigma}&= \frac{1}{\sqrt{2\pi \alpha}} e^{-\frac{i}{\sqrt{2}}[\phi_{\eta c}-\theta_{\eta c}+\sigma (\phi_{\eta s}- \theta_{\eta s})]}, \nn
L_{\eta \sigma}&= \frac{1}{\sqrt{2\pi \alpha}} e^{\frac{i}{\sqrt{2}}[\phi_{\eta c}+\theta_{\eta c}+\sigma (\phi_{\eta s}+ \theta_{\eta s})]} ,
\label{bos}
\end{align}
where $\alpha$ is the short-distance cut-off of the theory and we assume it to be the lattice constant of the NW. The bosonic fields satisfy the commutation relation $[\phi_{\eta \beta}(x),\theta_{\eta'\beta}(x')]= i\,\pi\,\delta_{\eta \eta'}\,\text{sgn}(x'-x)$. The field $\phi_{\eta  \beta}$ and $\theta_{\eta \beta}$ relate to the $\beta=c$ (charge) and $\beta=s$ (spin) density and current in the $\eta$-band, respectively. Thus we rewrite the linearized Hamiltonian in bosonized fields and obtain different terms in the Hamiltonian $H'$ defined in Eqs. (4)-(6) of the main text.  Notably, we also include interband interaction terms [see Eq. (4) of the main text]. In addition,  there is  a backscattering term involving spin up and spin down electrons for each Floquet band separately of the form
\begin{align}
H_g&= \sum_\eta  g_\eta \int dx \,R^\dagger_{\eta1} L_{\eta 1} L^\dagger_{\eta \bar 1} R_{\eta \bar 1} +\text{H.c.}= \sum_\eta\frac{g_\eta}{2\pi^2 \alpha^2} \int dx \cos(2\sqrt{2} \phi_{\eta s}),
\end{align}
where $g_{\eta}$ is the coupling strength. However, in the regime of $K_s \ge 1$, this term is either marginal or irrelevant [\onlinecite{MT2}]. Thus, we drop this term in the main text.

\section{Green functions of unperturbed Hamiltonian $H_{kin}$ in the presence of interband cross term} 
\label{appA}
In this Appendix, we derive first the Matsubara Green functions and later different correlation functions of $\phi-\theta$ fields in the presence of the cross term for the unperturbed Hamiltonian in $H'_{kin}$ defined in Eq. (4) of the main text.  The quadratic part of the Matsubara action is written as $S =S_c+S_s$.
Here we split the action into two parts $S_c$ and $S_s$, corresponding to the charge and spin-sectors, respectively, with the following form
\begin{align}
S_c &= \sum_{\eta = 1,\bar{1}}\frac{1}{\pi}\int dx\, d\tau \left[ i\partial_x \theta_{\eta c}\, \partial_\tau \phi_{\eta c} - u_{\eta c}\frac{(\partial_x \phi_{\eta c})^2 }{2K_{\eta c}} - \frac{u_{\eta c}K_{\eta c}}{2}(\partial_x \theta_{\eta c})^2\right] - \frac{\tilde{u}}{\pi} \int dx\, d\tau\;\left( \partial_x \phi_{1 c} \right) \left( \partial_x \phi_{\bar{1}c} \right), \nn
 S_s &= \sum_{\eta = 1,\bar{1}}\frac{1}{\pi}\int dx\, d\tau \left[ i\partial_x \theta_{\eta  s} \,\partial_\tau \phi_{\eta s} - u_{\eta c}\frac{(\partial_x \phi_{\eta s})^2 }{2K_{\eta s}}- \frac{u_{\eta s}K_{\eta s}}{2}(\partial_x \theta_{\eta s})^2\right].
\label{actionc}
\end{align}
The correlation functions  for the spin sector are unchanged as the cross-term between different Floquet band appears in the charge sector only and has the following form 
\begin{align}
&\langle [\phi_{\eta s}(x,\tau)- \phi_{\eta s} (0,0)]^2\rangle_0 = K_{\eta s} \,\text{ln}\frac{\sqrt{x^2+u_{\eta s}^2 \tau^2}}{\alpha}\,, \,\,\,\,\,\,\,\,\, \langle [\phi_{\eta s}(0,0)]^2\rangle_0 \sim -\frac{K_{\eta s}}{2} \,\text{ln} \alpha\, ,  \nonumber \\
&\langle [\theta_{\eta s}(x,\tau)- \theta_{\eta s}(0,0)]^2\rangle_0 =\frac{1}{K_{\eta s}} \,\text{ln}\frac{\sqrt{x^2+u_{\eta s}^2 \tau^2}}{\alpha}\,,\,\,\,\,\,\,\,\,\, \langle [\theta_{\eta s}(0,0)]^2\rangle_0 \sim -\frac{1}{2K_{\eta s}} \,\text{ln} \alpha\,,
\label{exp11} 
 \end{align}
where the expectation value $\langle \cdots\rangle_0$ is taken with respect to the  LL action $S_s$ defined in Eq. (\ref{actionc}). For the charge sector, we calculate the action in Fourier space which then becomes
\begin{align}
S_c=\frac{1}{2}\int \frac{dq\, d\omega}{4\pi^2} \,\Phi_c^\dagger(q,\omega) \check{G}_c^{-1}(q,\omega) \Phi_c(q,\omega),
\end{align}
where $\Phi_c(q,\omega)=[\phi_{1 c}(q,\omega),\ \theta_{1 c}(q,\omega),\ \phi_{\bar{1} c}(q,\omega),\ \theta_{\bar{1}c}(q,\omega)]$. Here, we use the definition $\phi_{\eta c}(q,\omega)= \int \, dx\, d \tau \,\phi_{\eta c}(x,\tau) \, e^{i\,(q\,x-\omega\,\tau)}$. The action $S_c$ [see Eq. (\ref{actionc})] determines the inverse Green function as 
\begin{align}
\check{G}_c^{-1}(q,\omega) = \frac{1}{\pi}
\begin{pmatrix}
q^2u_{1 c}K_{1 c}^{-1}& iq\omega & q^2\tilde{u}&0\\
iq\omega & q^2u_{1 c}K_{1 c}& 0& 0\\
q^2\tilde{u}&0& q^2 u_{\bar{1}c}K_{\bar{1}c}^{-1}&i q\omega\\
0&0&iq\omega & q^2u_{\bar{1}c}K_{\bar{1}c}
\end{pmatrix}.
\label{iGc}
\end{align} 
We calculate the inverse of Eq.  (\ref{iGc}) to obtain the matrix of Green functions for the charge sector which yields
\begin{align}
\check{G}_c(q,\omega) = \begin{pmatrix}
G_{\phi_{1 c} \phi_{1 c}} & G_{\phi_{1 c} \theta_{1 c}} & G_{\phi_{1 c} \phi_{\bar{1}c}} & G_{\phi_{1 c} \theta_{\bar{1}c}}\\
G_{\theta_{1 c} \phi_{1 c}} & G_{\theta_{1 c} \theta_{1 c}} & G_{\theta_{1 c} \phi_{\bar{1}c}} & G_{\theta_{1 c} \theta_{\bar{1}c}}\\
G_{\phi_{\bar{1}c} \phi_{1 c}} & G_{\phi_{\bar{1}c} \theta_{1 c}} & G_{\phi_{\bar{1}c} \phi_{\bar{1}c}} & G_{\phi_{\bar{1}c} \theta_{\bar{1}c}}\\
G_{\theta_{\bar{1}c} \phi_{1 c}} & G_{\theta_{\bar{1}c} \theta_{1 c}} & G_{\theta_{\bar{1}c} \phi_{\bar{1}c}} & G_{\theta_{\bar{1}c} \theta_{\bar{1}c}}\\
\end{pmatrix}.
\label{Gc}
\end{align}
For the calculations to follow below, we need to compute the correlations  $\langle \phi_{\eta c}\phi_{\eta c}\rangle_0 $ and $\langle \theta_{\eta c}\theta_{\eta c}\rangle_0$ in each of the $\eta$-NW and $\langle \phi_{\eta c}\phi_{\bar \eta c}\rangle_0$ and $\langle\theta_{\eta c}\theta_{\bar \eta c}\rangle_0$ in between $\eta-\bar \eta$ NWs. We write the corresponding Green functions as 
\begin{align}
&G_{\theta_{\eta c} \theta_{\eta c}}(\omega,q) = -\frac{\pi\,({K_{\eta c}} {K_{\bar{\eta}c}} q^{2} {\tilde{u}}^{2} {u_{\bar{\eta}c}} - q^{2} {u_{\eta c}} {u^{2}_{\bar{\eta}c}} - {u_{\eta c}}\, {\omega}^{2})}{ {\left[ (\omega^2+u_{\eta c}^2\, q^2)(\omega^2 + u_{\bar{\eta}c}^2 \,q^2)\right]} {K_{\eta c}}-{K^{2}_{\eta c}} {K_{\bar{\eta}c}} \,q^{4}\, {\tilde{u}}^{2} \,{u_{\eta c}} \,{u_{\bar{\eta}c}} },\nn
&G_{\theta_{\eta  c} \theta_{\bar{\eta}c}}(\omega,q) = \frac{\pi \,\tilde{u}\, \omega^2}{(\omega^2+u_{\eta c}^2 \,q^2)(\omega^2 + u_{\bar{\eta}c}^2 \,q^2) - K_{\eta  c}\,K_{\bar{\eta}c}\,q^4 \tilde{u}^2 \,u_{\eta c} \,u_{\bar{\eta}c}},\nn
&G_{\phi_{\eta  c} \phi_{\eta  c}}(\omega,q) =  \frac{ \pi\, K_{\eta c}\, u_{\eta c}\,(\omega^2 + q^2  \,u_{\bar{\eta}c}^2 ) }{(\omega^2+u_{\eta c}^2\, q^2)(\omega^2 + u_{\bar{\eta}c}^2 \,q^2) - K_{\eta  c}\,K_{\bar{\eta} c}\,q^4 \,\tilde{u}^2 \,u_{\eta c} \,u_{\bar{\eta}c}},\nn
&G_{\phi_{\eta  c} \phi_{\bar{\eta} c}}(\omega,q) =
-\frac{\pi \,K_{\eta  c}\, K_{\bar{\eta} c}\, q^2 \,u_{\eta c} \,u_{\bar{\eta}c} \tilde{u} }{(\omega^2+u_{\eta c}^2 \,q^2)(\omega^2 + u_{\bar{\eta}c}^2 \,q^2) - K_{\eta  c}\,K_{\bar{\eta}c}\,q^4 \,\tilde{u}^2 \,u_{\eta c} \,u_{\bar{\eta}c}}.
\end{align}
These expressions can be simplified in case of a band symmetry, $u_{\bar{1}c} = u_{1 c }= u_c$, and $K_{1 c} = K_{\bar{1}c} = K_c$, as follows
\begin{align}
&G_{\theta_{\eta c}\theta_{\eta c}}(\omega,q) = 
\frac{\pi}{2K_c}\left(\frac{ u_c+ \tilde{u} K_{c} }{\omega^2 + u_+^2 q^2} +
\frac{u_c-\tilde{u} K_c}{\omega^2 + u_-^2 q^2}\right),\nn
& G_{\theta_{\eta  c} \theta_{\bar{\eta}c}}(\omega,q) = \frac{\pi}{2 K_c}\left(
\frac{\tilde{u}\, K_c + u_c}{\omega^2 + u_+^2 q^2} + 
\frac{\tilde{u}\, K_c - u_c}{\omega^2 + u_-^2 q^2}\right),\nn
&G_{\phi_{\eta  c} \phi_{\eta  c}}(\omega,q) = 
\frac{K_c}{2} \left(\frac{\pi\, u_c}{\omega^2 + q^2\, u_+^2} + \frac{\pi\, u_c}{\omega^2 + q^2 u_-^2}\right), \nn
& G_{\phi_{\eta  c} \phi_{\bar{\eta} c}}(\omega,q) = \frac{\pi K_c \,u_c}{2}\left( \frac{1}{\omega^2 + q^2\, u_+^2 } - \frac{1}{\omega^2 + q^2 \,u_-^2 }   \right),
\end{align}
where we define $u_{\pm}^2 = u_c^2 \pm K_c u_c \, \tilde{u}$. If the cross-term is absent, $\tilde{u} = 0$, we get $u_\pm=u_c$ and recover the usual form for the correlation functions:
\begin{align}
G_{\theta_{\eta c} \theta_{\eta c}}(\omega,q) = \frac{1}{K_c} \frac{\pi u_c}{\omega^2 + u_c^2q^2},\;\;G_{\theta_{\eta c} \theta_{\bar \eta c}}(\omega,q) = 0,\nn
G_{\phi_{\eta c} \phi_{\eta c}}(\omega,q) = \frac{K_c}{2} \frac{\pi u_c}{\omega^2 + u_c^2q^2},\;\;G_{\phi_{\eta c} \phi_{\bar \eta c}}(\omega,q) = 0.
\label{A9}
\end{align}
In the time-coordinate representation,
${\pi u}/(\omega^2 + u^2q^2)$ leads to $F_1(x, u\tau) =\text{ln}\left(\sqrt{x^2+u^2 \tau^2}\,/\alpha\right)$ [\onlinecite{Giamarchi}]. Thus, for  the nonzero cross-term, $\tilde{u}\neq 0$, we get by analogy 
\begin{align}
& G_{\theta_{\eta c} \theta_{\eta c}} \equiv \frac{1}{2K_c} \frac{u_c + \tilde{u}K_c}{u_+}F_1(x, u_+\tau) + \frac{1}{2K_c} \frac{u_c - \tilde{u}K_c}{u_-}F_1(x, u_-\tau),\nn
& G_{\theta_{\eta  c} \theta_{\bar{\eta}c}} \equiv \frac{1}{2K_c} \frac{\tilde{u} K_c + u_c}{u_+} F_1(x,u_+ \tau) - \frac{1}{2K_c} \frac{u_c-\tilde{u} K_c }{u_-} F_1(x,u_- \tau),\nn
&G_{\phi_{\eta  c} \phi_{\eta  c}} \equiv \frac{K_c \,u_c}{2u_+} F_1(x,u_+\tau) +\frac{K_c \,u_c}{2u_-} F_1(x,u_-\tau), \nn
&  G_{\phi_{\eta  c} \phi_{\bar{\eta}c}} \equiv \frac{K_c\,u_c}{2\,u_+}F_1(x,u_+\tau) - 
\frac{K_c\,u_c}{2\,u_-}F_1(x,u_-\tau).
\label{A11}
\end{align}
Therefore, in the presence of the interband cross-term, the modified form of the correlation functions is given by
\begin{align}
&\langle [\theta_{\eta c}(x,\tau)- \theta_{\eta c}(0,0)]^2\rangle_0 =\lambda_+ \,\text{ln}\frac{\sqrt{x^2+u_+^2 \tau^2}}{\alpha}+ \lambda_- \,\text{ln}\frac{\sqrt{x^2+u_{-}^2 \tau^2}}{\alpha},\nn
&\langle [\theta_{\eta c}(x,\tau)- \theta_{\bar{\eta}c}(0,0)]^2\rangle_0 =\lambda_+ \,\text{ln}\frac{\sqrt{x^2+u_+^2 \tau^2}}{\alpha}- \lambda_- \,\text{ln}\frac{\sqrt{x^2+u_{-}^2 \tau^2}}{\alpha},\nn
&\langle [\phi_{\eta c}(x,\tau)- \phi_{{\eta}c}(0,0)]^2\rangle_0 = \frac{K_c \,u_c}{2\,u_+}\,\text{ln}\frac{\sqrt{x^2+u_+^2 \tau^2}}{\alpha}+ \frac{K_c \,u_c}{2\,u_-} \,\text{ln}\frac{\sqrt{x^2+u_{-}^2 \tau^2}}{\alpha},\nn
&\langle [\phi_{\eta c}(x,\tau)- \phi_{\bar{\eta}c}(0,0)]^2\rangle_0 = \frac{K_c \,u_c}{2\,u_+}\,\text{ln}\frac{\sqrt{x^2+u_+^2 \tau^2}}{\alpha}- \frac{K_c \,u_c}{2\,u_-} \,\text{ln}\frac{\sqrt{x^2+u_{-}^2 \tau^2}}{\alpha},
\label{A12}
 \end{align}
 where we define, $\lambda_{\pm}=\frac{1}{2K_c} \frac{u_c \pm \tilde{u}K_c}{u_\pm}$. We will utilize these correlation functions in the next section to obtain the RG flow equations for different system parameters.  

 \section{RG equations for coupling constants, velocities, and LL parameters}
 \label{appB}

\subsection{RG equations for $H'_{sc}$}
\label{ARG1}
Below we show the calculation of RG equations for two terms, namely $H'_{sc}$ and $H'_Z$. We define complex coordinates $(z_{\kappa\nu},\bar z_{\kappa\nu})$ as $z_{\kappa\nu}=-i \,x_\kappa+u_\nu \tau_\kappa$ and $\bar z_{\kappa\nu}=i \,x_\kappa+u_\nu \tau_\kappa$, where $\nu \in \{c,s,+,-\}$ is the index for different velocities and $\kappa$ is the index for position-time coordinates. We also define the center-of-mass coordinate for charge-spin sector as, $(z_{1\nu}, z_{2\nu}) \Rightarrow (\chi_{\nu}, \zeta_{\nu})$ where $\chi_{\nu}=(z_{1\nu}-  z_{2\nu})$ and $\zeta_{\nu}=(z_{1\nu}+  z_{2\nu})/2$, and $(\bar z_{1\nu}, \bar z_{2\nu}) \Rightarrow (\bar \chi_{\nu}, \bar \zeta_{\nu})$ where $\bar \chi_{\nu}=(\bar z_{1\nu}-  \bar z_{2\nu})$ and $\bar \zeta_{\nu}=(\bar z_{1\nu}+ \bar z_{2\nu})/2$.

First, we focus on the superconducting term $H'_{sc}= \Delta_{sc} \sum_\eta\left(\cos [\sqrt{2}\,(\theta_{\eta c}+\phi_{\eta s})]+\cos [\sqrt{2}\,(\theta_{\eta c}-\phi_{\eta s})]\right)/\pi \, \alpha$ and obtain the RG flow equations for  a dimensionless coupling constant $\tilde \Delta_{sc}= \Delta_{sc} \,\alpha/u_c $. Also, below  we show the calculation just for the  first cosine term, however, we obtain the same corrections from the second cosine term. We make use of the operator product expansion from Refs. [\onlinecite{Cardi},\onlinecite{MT2}] and expanded the partition function in powers of the cosine term characterized by a real constant $\lambda$ up to the second-order 
\begin{align}
&Z_a= Z_{0a} \,\Big\langle\, 1-  \sum_\eta \frac{\tilde\Delta_{sc} u_{c}}{ \pi \, \alpha^2}\int dx \,d\tau\ \cos\big(\lambda [\theta_{\eta c}(x,\tau)+\phi_{\eta s}(x,\tau)]\big) \label{7} \\
 &\hspace*{0.8cm}+ \frac{(\tilde \Delta_{sc})^2\, u_c^2}{2\, \pi^2 \, \alpha^4}\int dx_1\, dx_2 \,d\tau_1 \,d\tau_2\, \cos\big(\lambda \,[\theta_{\eta c}(x_1,\tau_1)+\phi_{\eta s}(x_1,\tau_1)]\big) \cos\big(\lambda \,[\theta_{\eta c}(x_2,\tau_2)+\phi_{\eta s}(x_2,\tau_2)]\big)  + \cdots  \Big\rangle_0 \,.
\nonumber
\end{align} 
The second term has the following form in complex coordinates 
\begin{align}
&\cos\big(\lambda\,[\theta_{\eta c}(z_{1c},\bar z_{1c})+\phi_{\eta s}(z_{1s},\bar z_{1s})]\big)\, \cos\big(\lambda\,[\theta_{\eta c}(z_{2c},\bar z_{2c})+\phi_{\eta s}(z_{2s},\bar z_{2s})]\big)\label{8} \\
&\simeq (e^{ i\,\lambda[\theta_{\eta c}(z_{1c},\bar z_{1c})+\phi_{\eta s}(z_{1s},\bar z_{1s})] } e^{ -i\,\lambda[\theta_{\eta c}(z_{2c},\bar z_{2c})+\phi_{\eta s}(z_{2s},\bar z_{2s})] }+e^{ -i\,\lambda[\theta_{\eta c}(z_{1c},\bar z_{1c})+\phi_{\eta s}(z_{1s},\bar z_{1s})] } e^{ i\,\lambda[\theta_{\eta c}(z_{2c},\bar z_{2c})+\phi_{\eta s}(z_{2s},\bar z_{2s})] })/4 \,\nn
&\simeq\frac{1}{2(|\chi_{+}|/\alpha)^{ \lambda^2\lambda_+/2} (|\chi_-|/\alpha)^{ \lambda^2\lambda_-/2}(|\chi_s|/\alpha)^{\lambda^2 K_s/2}}\nn
& \times\Bigg( 1- \lambda^2  |\chi_c|^2 \,[J_{\theta_{\eta c}} \bar J_{\theta_{\eta c}}]_{(\zeta_c,\bar \zeta_c)} - \lambda^2  |\chi_s|^2 \,[J_{\phi_{\eta s}} \bar J_{\phi_{\eta s}}]_{(\zeta_s,\bar \zeta_s)} \nn
&+ 
\frac{\lambda^2}{2} \Big[ (\chi_c)^2 [J^2_{\theta_{\eta c}}]_{(\zeta_c,\bar \zeta_c)}+ (\bar \chi_c)^2  [\bar J^2_{\theta_{\eta c}}]_{(\zeta_c,\bar \zeta_c)} + (\chi_s)^2 [J^2_{\phi_{\eta s}}]_{(\zeta_s,\bar \zeta_s)}+ (\bar \chi_s)^2  [\bar J^2_{\phi_{\eta s}}]_{(\zeta_s,\bar \zeta_s)} \Big] +\cdots \Bigg), \nonumber
\end{align}
where $J_{\phi_{\eta \beta}}=i\partial_{\zeta_\beta} \phi_{\eta \beta}$, $\bar J_{\phi_{\eta \beta}}= -i \partial_{\zeta_\beta} \phi_{\eta \beta}$, $J_{\theta_{\eta \beta}}=i\partial_{\zeta_\beta} \theta_{\eta \beta}$ and $\bar J_{\theta_{\eta \beta}}= -i \partial_{\zeta_\beta} \theta_{\eta \beta}$ with $\beta=c/s$. For the case of identical Floquet bands, we omit the index $\eta$ such that $K_{\eta \beta}= K_{\beta}$ and $u_{\eta \beta}=u_{\beta}$. We also use the relation between two bosonic operators $e^A e^B= :e^{A+B}: e^{\langle AB+\frac{A^2+B^2}{2}\rangle_0}$. The expectation value is taken with respect to the Hamiltonian $H'_{kin}$. Further, we utilize the correlation functions described in Appendix A by rewriting them in terms of complex coordinate as 
\begin{align}
&\langle [\theta_{\eta c}(z_{1c},\bar z_{1c})- \theta_{\eta c}(z_{2c},\bar z_{2c})]^2\rangle_0 =\lambda_+ \,\text{ln}\frac{|\chi_{+}|}{\alpha}+ \lambda_- \,\text{ln}\frac{|\chi_{-}|}{\alpha},\nonumber\\
&\langle [\theta_{\eta c}(z_{1c},\bar z_{1c})- \theta_{\bar{\eta}c}(z_{2c},\bar z_{2c})]^2\rangle_0 =\lambda_+ \,\text{ln}\frac{|\chi_{+}|}{\alpha}- \lambda_- \,\text{ln}\frac{|\chi_{-}|}{\alpha},\nonumber\\
&\langle [\phi_{\eta c}(z_{1c},\bar z_{1c})- \phi_{{\eta}c}(z_{2c},\bar z_{2c})]^2\rangle_0 = \frac{K_c \,u_c}{2\,u_+}\,\text{ln}\frac{|\chi_{+}|}{\alpha}+ \frac{K_c \,u_c}{2\,u_-} \,\text{ln}\frac{|\chi_{-}|}{\alpha},\nonumber\\
&\langle [\phi_{\eta c}(z_{1c},\bar z_{1c})- \phi_{\bar{\eta}c}(z_{2c},\bar z_{2c})]^2\rangle_0 = \frac{K_c \,u_c}{2\,u_+}\,\text{ln}\frac{|\chi_{+}|}{\alpha}- \frac{K_c \,u_c}{2\,u_-} \,\text{ln}\frac{|\chi_{-}|}{\alpha},\nonumber\\
&\langle [\phi_{\eta s}(z_{1s},\bar z_{1s})- \phi_{\eta s} (z_{2s},\bar z_{2s})]^2\rangle_0 = K_s \,\text{ln}\frac{|\chi_{s}|}{\alpha}\,, \,\,\,\,\,\,\,\,\, \langle [\phi_{\eta \beta}(z_{\kappa\beta}, \bar z_{\kappa\beta})]^2\rangle_0\sim\langle \phi_{\eta \beta}(z_{\kappa\beta}, \bar z_{\kappa\beta})\phi_{\bar \eta \beta}(z_{\kappa\beta}, \bar z_{\kappa\beta})\rangle_0 \propto -\frac{K_{\beta}}{2} \,\text{ln} \,\alpha\, ,  \nonumber \\
&\langle [\theta_{\eta  s}(z_{1s},\bar z_{1s})- \theta_{\eta  s}(z_{2s},\bar z_{2s})]^2\rangle_0 =\frac{1}{K_s} \,\text{ln}\frac{|\chi_{s}|}{\alpha}\,,\,\,\,\,\,\,\,\,\,\, \langle [\theta_{\eta \beta}(z_{\kappa\beta}, \bar z_{\kappa\beta})]^2\rangle_0\sim\langle \theta_{\eta \beta}(z_{\kappa\beta}, \bar z_{\kappa\beta})\theta_{\bar \eta \beta}(z_{\kappa\beta}, \bar z_{\kappa\beta})\rangle_0 \propto -\frac{1}{2K_{\beta}} \,\text{ln}    
 \,\alpha\,.
 \label{cfz}
\end{align}
We also define $\partial_{\zeta_\beta}= -\frac{1}{2} \left(\frac{\partial_\mathcal{T}}{u_\beta}-i \, \partial_X \right)$ and $\partial_{ \bar {\zeta_\beta}}= -\frac{1}{2} \left(\frac{\partial_\mathcal{T}}{u_\beta}+i\,\partial_X \right)$  as well as
\begin{align}
& J_{\phi_{\eta\beta}} \bar J_{\phi_{\eta\beta}}= \partial_{\zeta_\beta} \phi_{\eta\beta} \, \partial_{\bar \zeta_\beta}\phi_{\eta\beta}= \frac{(\partial_X \phi_{\eta\beta})^2+(\partial_\mathcal{T} \phi_{\eta\beta})^2/u_\beta^2}{4}, \nn 
&J_{\phi_{\eta\beta}}^2= -(\partial_{\zeta_\beta} {\phi_{\eta\beta}})^2= \frac{(\partial_X \phi_{\eta\beta})^2-(\partial_\mathcal{T} \phi_{\eta\beta})^2/u_\beta^2+ 2\,i\,(\partial_X \phi_{\eta\beta}) (\partial_\mathcal{T} \phi_{\eta\beta})/u_\beta }{4},\nn
&\bar J_{\phi_{\eta\beta}}^2= -(\partial_{ \bar \zeta_\beta}\phi_{\eta\beta})^2= \frac{(\partial_X \phi_{\eta\beta})^2-(\partial_\mathcal{T} \phi_{\eta\beta})^2/u_\beta^2- 2\,i\,(\partial_X \phi_{\eta\beta}) (\partial_\mathcal{T} \phi_{\eta\beta})/u_\beta }{4}.
\label{kineticterms}
\end{align}
We can write the same relations for the $\theta$-field. Next, for $\lambda=\sqrt{2}$, the scaling dimension of the second term in Eq. (\ref{7}) term is $(\lambda_+ +\lambda_-+ K_s)/2$. Thus, the RG equation is given by $d\tilde \Delta_{sc}/dl= [2-(\lambda_+ +\lambda_-+ K_s)/2]\, \tilde \Delta_{sc}$. We can go back to $(x,\tau,X,\mathcal{T})$ center-of-mass notation and utilize the fact that for the odd terms the integral over $dx\, d\tau$ gives zero and hence calculate the terms contributing to the kinetic energy in charge sector from Eq. (\ref{8}) as following
\begin{align}
- \lambda^2 |\chi_c|^2 \,[J_{\theta_{\eta c}} \bar J_{\theta_{\eta c}}]_{(\zeta_c,\bar \zeta_c)} +\frac{\lambda^2}{2}  (\chi_c)^2 [J^2_{\theta_{\eta c}}]_{(\zeta_c,\bar \zeta_c)}+ (\bar \chi_c)^2  [\bar J^2_{\theta_{\eta c}}]_{(\zeta_c,\bar \zeta_c)}= -\frac{\lambda^2 }{2} \Big[x^2\,(\partial_X \theta_{\eta c})^2 + \tau^2 (\partial_\mathcal{T} \theta_{\eta c})^2 \Big].
\end{align}
The second-order term in the $Z_{a}$ for the charge sector has the form
\begin{align}
I_c&= -\sum_\eta \frac{(\tilde\Delta_{sc})^2 \,u_c^2\,\lambda^2}{8\,\pi^2\, \alpha^4}  \int dX d\mathcal T \bigg[(\partial_X \theta_{\eta c})^2 \int \frac{x^2\,dx \,d\tau}{(|\chi_+|/\alpha)^{ \lambda^2\lambda_+/2} (|\chi_-|/\alpha)^{ \lambda^2\lambda_-/2}(|\chi_s|/\alpha)^{\lambda^2 K_s/2}} \nn
&\hspace*{3.5cm}+ (\partial_{\mathcal T} \theta_{\eta c})^2 \int \frac{\tau^2\,dx \,d\tau}{(|\chi_+|/\alpha)^{ \lambda^2\lambda_+/2} (|\chi_-|/\alpha)^{ \lambda^2\lambda_-/2}(|\chi_s|/\alpha)^{\lambda^2 K_s/2}}\bigg].
\label{9}
\end{align}
We change to polar coordinates such that $(x,u_c \tau)$ is $[r \cos\theta', r \sin\theta']$ such that $\int dx\,(u_c\,d\tau)=\int r\, dr \,d\theta'$ and calculate the integral over $r$ for the range $[\alpha,\alpha+d\alpha]$. 
\begin{align}
I_c&= -\sum_\eta \frac{(\tilde\Delta_{sc})^2 \,u_c\,\lambda^2}{2}\left( \frac{d\alpha}{\alpha}\right) \int \frac{ dX d\mathcal T}{2\,\pi} \bigg[(\partial_X \theta_{\eta c})^2 h_x^\Delta + \frac{(\partial_{\mathcal T} \theta_{\eta c})^2}{u_c^2} h_\tau^\Delta\bigg] ,
\label{Ic}
\end{align}
where we define
\begin{align}
2\,\pi\,h_x^\Delta=& \int \frac{ \cos^2\theta' \,d\theta'  }{\left(\cos^2\theta'+\frac{u_+^2}{u_c^2} \sin^2\theta'\right)^{ \lambda^2\lambda_+/4 } \left(\cos^2\theta'+\frac{u_-^2}{u_c^2} \sin^2(\theta')\right)^{ \lambda^2\lambda_-/4}\left(\cos^2\theta'+\frac{u_s^2}{u_c^2} \sin^2\theta'\right)^{\lambda^2 K_s/4}}\,,\nn
2\,\pi\,h_\tau^\Delta=&\int \frac{\sin^2\theta' \,d\theta'  }{\left(\cos^2\theta'+\frac{u_+^2}{u_c^2} \sin^2\theta'\right)^{ \lambda^2\lambda_+/4} \left(\cos^2\theta'+\frac{u_-^2}{u_c^2} \sin^2\theta'\right)^{ \lambda^2\lambda_-/4}\left(\cos^2\theta'+\frac{u_s^2}{u_c^2} \sin^2\theta'\right)^{\lambda^2 K_s/4}}.
\end{align}
From the equation of motion, we get $\partial_{\mathcal T} \theta_{\eta c}=-i [\frac{u_c}{K_c}\partial_X \phi_{\eta c}+ \tilde u\, \partial_X \phi_{\bar \eta c}]$. Therefore for $\lambda=\sqrt{2}$, Eq. (\ref{Ic}) takes following form
\begin{align}
I_c&= -(\tilde\Delta_{sc})^2 \,u_c\left( \frac{d\alpha}{\alpha}\right) \int \frac{ dX d\mathcal T}{2\,\pi} \bigg[(\partial_X \theta_{1 c})^2 h_x^\Delta +(\partial_X \theta_{\bar 1 c})^2 h_x^\Delta - \frac{4\,  \tilde u\,(\partial_X \phi_{ 1 c})\,(\partial_X \phi_{\bar 1 c})}{u_c\,K_c} h_\tau^\Delta \nn
&\hspace*{7.cm}- \frac{(u_c^2/K_c^2+\tilde u^2)(\partial_X \phi_{ 1 c})^2+(u_c^2/K_c^2+\tilde u^2)(\partial_X \phi_{\bar 1 c})^2}{u_c^2} h_\tau^\Delta
 \bigg],
\end{align}
Comparing with Eq. (4) in the main text, we obtain the RG equation for the charge sector as
\begin{align}
&\frac{d}{dl} \left(\frac{u_c}{K_c}\right)= -(\tilde \Delta_{sc})^2\,\left( \frac{u_c}{K_c^2}+ \frac{\tilde u^2}{u_c}\right)\, h_\tau^\Delta, \,\,\frac{d}{dl} \left(u_c\,K_c\right)= (\tilde \Delta_{sc})^2\, u_c\,h_x^\Delta,\,\, \frac{d\tilde u}{dl} =  \frac{-2\,\tilde u}{K_c}(\tilde \Delta_{sc})^2\, h_\tau^\Delta, \label{u1}\\[1em]
&\frac{du_c}{dl}= \frac{(\tilde \Delta_{sc})^2}{2} 
\left[\frac{u_c}{K_c}(h_x^\Delta-h_\tau^\Delta)-\frac{h_\tau^\Delta\,K_c\,\tilde u^2}{u_c}\right], \, \, \frac{dK_c}{dl}=\frac{(\tilde \Delta_{sc})^2}{2} 
\left[(h_x^\Delta+h_\tau^\Delta)+\frac{h_\tau^\Delta\,K_c^2\,\tilde u^2}{u_c^2}\right].
\label{11}
\end{align}
For the spin sector (corresponding to the field $\phi_{\eta s}$), we get a similar expression like Eq. (\ref{Ic}), however, $\theta_{\eta c}$ gets replaced by $\phi_{\eta s}$,
\begin{align}
I_s= -\frac{(\tilde\Delta_{sc})^2 \,u_c\,\lambda^2}{2}\left( \frac{d\alpha}{\alpha}\right) \int \frac{ dX d\mathcal T}{2\,\pi} \bigg[(\partial_X \phi_{\eta s})^2 h_x^\Delta + \frac{(\partial_{\mathcal T} \phi_{\eta s})^2}{u_c^2} h_\tau^\Delta\bigg].
\label{12}
 \end{align}
We use $\partial_{\mathcal T} \phi_{\eta s}=-i\,u_{s}\, K_s\,\partial_X \theta_{\eta s}$ and obtain
\begin{align}
I_s= -\frac{(\tilde\Delta_{sc})^2 \,u_c\,\lambda^2}{2}\left( \frac{d\alpha}{\alpha}\right) \int \frac{ dX d\mathcal T}{2\,\pi} \bigg[(\partial_X \phi_{\eta s})^2 h_x^\Delta - \frac{u_s^2\,K_s^2(\partial_X \theta_{\eta s})^2}{u_c^2} h_\tau^\Delta\bigg].
\label{13}
 \end{align}
As a result, the RG equations become
\begin{align}
&\frac{d}{dl} \left(\frac{u_s}{K_s}\right)= (\tilde \Delta_{sc})^2 \,u_c\, h_x^\Delta, \,\,\frac{d}{dl} \left(u_s\,K_s\right)= -\frac{u_s^2\,K_s^2(\tilde \Delta_{sc})^2\,  h_\tau^\Delta}{u_c}, \nn[1em]
&\frac{du_s}{dl}=\frac{(\tilde \Delta_{sc})^2\,u_c\,K_s }{2}\left(h_x^\Delta-\frac{u_s^2}{u_c^2} \,h_\tau^\Delta\right), \, \, \frac{dK_s}{dl}=-\frac{(\tilde \Delta_{sc})^2 \,K_s^2}{2}\left(\frac{u_c}{u_s}\,h_x^\Delta+\frac{u_s}{u_c} \,h_\tau^\Delta\right).
\label{14}
 \end{align}
We also note that the RG equations for $\cos [\sqrt{2}\,(\theta_{\eta c}-\phi_{\eta s})]/\pi \, \alpha$ have exactly the same form as written in Eqs. (\ref{11}) and (\ref{14}). Thus, in the final RG equations, we double the corrections and obtain the final contribution from $\tilde \Delta_{sc}$ as
\begin{align}
 &\frac{d\tilde u}{dl} =  \frac{-4\,\tilde u}{K_c}(\tilde \Delta_{sc})^2\, h_\tau^\Delta,\,\,\, \frac{du_c}{dl}= (\tilde \Delta_{sc})^2
\left[\frac{u_c}{K_c}(h_x^\Delta-h_\tau^\Delta)-\frac{h_\tau^\Delta\,K_c\,\tilde u^2}{u_c}\right], \, \, \, \frac{dK_c}{dl}=(\tilde \Delta_{sc})^2
\left[(h_x^\Delta+h_\tau^\Delta)+\frac{h_\tau^\Delta\,K_c^2\,\tilde u^2}{u_c^2}\right],\nn[1em]
&\frac{du_s}{dl}=(\tilde \Delta_{sc})^2\,u_c\,K_s \left(h_x^\Delta-\frac{u_s^2}{u_c^2} \,h_\tau^\Delta\right), \, \, \frac{dK_s}{dl}=-(\tilde \Delta_{sc})^2 \,K_s^2\left(\frac{u_c}{u_s}\,h_x^\Delta+\frac{u_s}{u_c} \,h_\tau^\Delta\right).
 \end{align}

\subsection{RG equations for $H'_Z$}
\label{ARG2}
In the following section, we calculate the RG equations for the Floquet Zeeman term $H'_Z$ given by 
\begin{align}
H'_Z=\sum_\eta \frac{t_F}{\pi \, \alpha} \cos \Big[\frac{1}{\sqrt{2}}\,(\phi_{\eta c}-\theta_{\eta c}-\phi_{\eta s}+\theta_{\eta s}+\phi_{\bar \eta c}+\theta_{\bar \eta c}+\phi_{\bar \eta s}+\theta_{\bar\eta s})\Big].
\label{15}
\end{align} 
Further, introducing $\tilde t_F=t_F\, \alpha/u_c$ as a dimensionless coupling constant, $\tilde \varphi_\eta=\phi_{\eta c}-\theta_{\eta c}-\phi_{\eta s}+\theta_{\eta s}+\phi_{\bar \eta c}+\theta_{\bar \eta c}+\phi_{\bar\eta s}+\theta_{\bar\eta s}$ and using the definition given in Appendix \ref{ARG1}, we write the correlator $\langle[\tilde \varphi_\eta(x_1,t_1) -\tilde \varphi_\eta(x_2,t_2)]^2 \rangle$ as
\begin{align}
&\langle[\tilde \varphi_\eta(x_1,t_1) -\tilde \varphi_\eta(x_2,t_2)]^2 \rangle \sim \langle[ \phi_{\eta c}(z_{1c},\bar z_{1c}) -\phi_{\eta c}(z_{2c},\bar z_{2c})]^2 \rangle+ \langle[ \theta_{\eta c}(z_{1c},\bar z_{1c}) -\theta_{\eta c}(z_{2c},\bar z_{2c})]^2 \rangle\nonumber\\ &+ \langle[ \phi_{\bar \eta c}(z_{1c},\bar z_{1c}) -\phi_{\bar \eta c}(z_{2c},\bar z_{2c})]^2 \rangle 
+ \langle[ \theta_{\bar \eta c}(z_{1c},\bar z_{1c}) -\theta_{\bar \eta c}(z_{2c},\bar z_{2c})]^2 \rangle+ \langle[ \phi_{\eta s}(z_{1s},\bar z_{1s}) -\phi_{\eta s}(z_{2s},\bar z_{2s})]^2 \rangle
\nn &+ \langle[ \theta_{\eta s}(z_{1s},\bar z_{1s}) -\theta_{\eta s}(z_{2s},\bar z_{2s})]^2 \rangle  +\langle[ \phi_{\bar \eta s}(z_{1s},\bar z_{1s}) -\phi_{\bar \eta s}(z_{2s},\bar z_{2s})]^2 \rangle 
+ \langle[ \theta_{\bar \eta s}(z_{1s},\bar z_{1s}) -\theta_{\bar \eta s}(z_{2s},\bar z_{2s})]^2 \rangle
\nn &-2\,\langle\phi_{\eta c}(z_{1c},\bar z_{1c}) \phi_{\bar \eta c}(z_{2c},\bar z_{2c})\rangle - 2\,\langle\phi_{\bar \eta c}(z_{1c},\bar z_{1c}) \phi_{\eta c}(z_{2c},\bar z_{2c})\rangle+2\,\langle\phi_{\eta c}(z_{1c},\bar z_{1c}) \phi_{\bar \eta c}(z_{1c},\bar z_{1c})\rangle
+ 2\,\langle\phi_{ \eta c}(z_{2c},\bar z_{2c}) \phi_{\bar \eta c}(z_{2c},\bar z_{2c})\rangle
\nn&+2\,\langle\theta_{\eta c}(z_{1c},\bar z_{1c}) \theta_{\bar \eta c}(z_{2c},\bar z_{2c})\rangle+ 2\,\langle\theta_{\bar \eta c}(z_{1c},\bar z_{1c}) \theta_{\eta c}(z_{2c},\bar z_{2c})\rangle-2\,\langle\theta_{\eta c}(z_{1c},\bar z_{1c}) \theta_{\bar \eta c}(z_{1c},\bar z_{1c})\rangle- 2\,\langle\theta_{ \eta c}(z_{2c},\bar z_{2c}) \theta_{\bar \eta c}(z_{2c},\bar z_{2c})\rangle.
\end{align}
Here we neglect the cross terms between $\theta$- and $\phi$-fields. After utilizing the correlation functions written in Eq. (\ref{cfz}) and considering identical bands ($K_{\eta \beta}= K_{\bar \eta \beta}= K_{\beta}$ and $u_{\eta \beta}= u_{\bar \eta \beta}= u_{\beta }$) with $\beta=c,s$, we get the following form 
\begin{align}
&\langle[\tilde \varphi_\eta(x_1,t_1) -\tilde \varphi_\eta(x_2,t_2)]^2 \rangle= 2 \,\bigg[\frac{K_c \,u_c}{2\,u_+}\,\text{ln}\frac{|\chi_{+}|}{\alpha}+ \frac{K_c \,u_c}{2\,u_-} \,\text{ln}\frac{|\chi_{-}|}{\alpha}+ \lambda_+ \,\text{ln}\frac{|\chi_{+}|}{\alpha}+ \lambda_- \,\text{ln}\frac{|\chi_{-}|}{\alpha} +K_s \,\text{ln}\frac{|\chi_{s}|}{\alpha}+\frac{1}{K_s} \,\text{ln}\frac{|\chi_{s}|}{\alpha}\nonumber\\
&+ \frac{K_c \,u_c}{2\,u_+}\,\text{ln}\frac{|\chi_{+}|}{\alpha}- \frac{K_c \,u_c}{2\,u_-} \,\text{ln}\frac{|\chi_{-}|}{\alpha}- \lambda_+ \,\text{ln}\frac{|\chi_{+}|}{\alpha}+ \lambda_- \,\text{ln}\frac{|\chi_{-}|}{\alpha}\bigg]
= 4\, \nu_+\,\text{ln}\frac{|\chi_{+}|}{\alpha}+4\,\lambda_- \,\text{ln}\frac{|\chi_{-}|}{\alpha}+2\,\nu_s\,\text{ln}\frac{|\chi_{s}|}{\alpha}.
\label{varphi}
\end{align}
Here we define $\nu_+= K_c u_c/(2\,u_+)$ and $\nu_s= K_s+ 1/ K_s$. We utilize the correlation function written in Eq. (\ref{varphi}) and obtain the most singular term in the OPEs for $\tilde t_F$ as
\begin{align}
&\cos[\lambda\tilde \varphi_1(x_1,t_1)]\cos[\lambda\tilde \varphi_1(x_2,t_2)]= \nn
&= \frac{1}{2(|\chi_+|/\alpha)^{ 2\,\lambda^2\nu_+} (|\chi_-|/\alpha)^{ 2\,\lambda^2\lambda_-}(|\chi_s|/\alpha)^{\lambda^2 \nu_s}} \times\Bigg( 1-\lambda^2  |\chi_c|^2 \,[J_{\phi_{1c}} \bar J_{\phi_{1c}}+ J_{\theta_{1c}} \bar J_{\theta_{1c}}+ J_{\phi_{\bar 1c}} \bar J_{\phi_{\bar 1c}}+ J_{\theta_{\bar 1c}} \bar J_{\theta_{\bar 1c}}]_{(\zeta_c,\bar \zeta_c)}\nn
& - \lambda^2 |\chi_s|^2 \,[J_{\phi_{1s}} \bar J_{\phi_{1s}}+ J_{\theta_{1s}} \bar J_{\theta_{1s}}+J_{\phi_{\bar 1s}} \bar J_{\phi_{\bar 1s}} + J_{\theta_{\bar 1s}} \bar J_{\theta_{\bar 1s}}]_{(\zeta_s,\bar \zeta_s)} -\lambda^2 |\chi_c|^2 [J_{\phi_{1c}} \bar J_{\phi_{\bar 1c}}+J_{\phi_{\bar 1c}} \bar J_{\phi_{ 1c}}- J_{\theta_{1c}} \bar J_{\theta_{\bar 1c}}- J_{\theta_{\bar 1c}} \bar J_{\theta_{ 1c}}]_{(\zeta_c,\bar \zeta_c)}\nn
&+ \frac{\lambda^2}{2} \Big[ (\chi_c)^2\, [J^2_{\phi_{1c}}+J^2_{\theta_{1c}}+J^2_{\phi_{\bar 1c}}+J^2_{\theta_{\bar 1c}}]_{(\zeta_c,\bar \zeta_c)}+ (\bar \chi_c)^2\,  [\bar J^2_{\phi_{1c}}+\bar J^2_{\theta_{1c}}+\bar J^2_{\phi_{\bar 1c}}+\bar J^2_{\theta_{\bar 1c}}]_{(\zeta_c,\bar \zeta_c)}  \Big] \nn 
&+ \lambda^2\bigg[(\chi_c)^2 \,[J_{\phi_{1c}} J_{\phi_{\bar 1c}}]_{(\zeta_c,\bar \zeta_c)} + (\bar \chi_c)^2\, [\bar J_{\phi_{1c}} \bar J_{\phi_{\bar 1c}}]_{(\zeta_c,\bar \zeta_c)}-(\chi_c)^2 \,[J_{\theta_{1c}} J_{\theta_{\bar 1c}}]_{(\zeta_c,\bar \zeta_c)}  -(\bar \chi_c)^2 [\bar J_{\theta_{1c}} \bar J_{\theta_{\bar 1c}}]_{(\zeta_c,\bar \zeta_c)}\bigg]\nn
&+\frac{\lambda^2}{2}\bigg[(\bar \chi_s)^2\,  [\bar J^2_{\phi_{1s}}+\bar J^2_{\theta_{1s}}+\bar J^2_{\phi_{\bar 1s}}+\bar J^2_{\theta_{\bar 1s}}]_{(\zeta_s,\bar \zeta_s)} + (\chi_s)^2\, [J^2_{\phi_{1s}}+J^2_{\theta_{1s}}+J^2_{\phi_{\bar 1s}}+J^2_{\phi_{\bar 1s}}]_{(\zeta_s,\bar \zeta_s)}\bigg]\Bigg)+\cdots.
\label{16}
\end{align}
To obtain the RG flow equations, we expand the partition function up to second-order in the cosine term,
\begin{align}
Z=Z_0\, \Big \langle 1-\sum_\eta \bigg(\, \frac{\tilde t_F\,u_c}{\pi\, \alpha^2} 
&\int dx\, d\tau\,\cos\left[\frac{1}{\sqrt{2}} \tilde \varphi(x,\tau)\right]\nn
&+ \frac{(\tilde t_F)^2\,u_c^2}{2\,\pi^2\, \alpha^4} \int dx_1\, d\tau_1 \,dx_2\, d\tau_2\cos\left[\frac{1}{\sqrt{2}} \tilde \varphi(x_1, \tau_1)\right]\cos\left[\frac{1}{\sqrt{2}} \tilde \varphi(x_2, \tau_2)\right]+ \cdots\bigg) \Big\rangle.
\label{17}
\end{align}
Thus, from Eq. (\ref{16}), by scaling-dimension arguments, the RG flow of $\tilde t_F$  is given by
\begin{align}
\frac{d\tilde t_F}{dl}= \left[2-\frac{( 2\,\nu_++ 2\,\lambda_- + \nu_s )}{4}\right]\, \tilde t_F,
\end{align}
where we take $\lambda=1/\sqrt{2}$ and $dl= d\alpha/\alpha$. Below, we calculate the second order correction from $\tilde t_F$ to the charge-spin LL parameters and velocities.

\subsubsection{Terms contributing to the kinetic energy}
We consider the third term in Eq. (\ref{17}) and collect all the terms contributing to the kinetic energy for the charge sector in the $\eta=1$  NW from Eq. (\ref{16}) and find
\begin{align}
I^{d\alpha}=& \frac{(\tilde t_F)^2\,u_c^2}{2\,\pi^2\, \alpha^4}\sum_\eta\int dx \,d\tau\, dX\, d\mathcal{T} \frac{1}{2(|\chi_+|/\alpha)^{2\, \lambda^2\nu_+} (|\chi_-|/\alpha)^{ 2\,\lambda^2\lambda_-}(|\chi_s|/\alpha)^{\lambda^2 \nu_s}} \bigg(-\lambda^2  |\chi_c|^2 \,\left[J_{\phi_{\eta c}} \bar J_{\phi_{\eta c}}+ J_{\theta_{\eta c}} \bar J_{\theta_{\eta c}}\right]_{(\zeta_c,\bar \zeta_c)}\nn
&\hspace*{5cm}+ \frac{\lambda^2\,(\chi_c)^2}{2} \, \left[J^2_{\phi_{\eta c}}+J^2_{\theta_{\eta c}}\right]_{(\zeta_c,\bar \zeta_c)}+\frac{\lambda^2\,(\bar \chi_c)^2}{2}\,  \left[\bar J^2_{\phi_{\eta c}}+\bar J^2_{\theta_{\eta c}}\right]_{(\zeta_c,\bar \zeta_c)}
\bigg),
\label{19}
\end{align} 
where the terms containing $J_{\phi_{\eta c}/\theta_{\eta c}}$ and $\bar J_{\phi_{\eta c}/\theta_{\eta c}}$ take the form 
\begin{align}
&-\lambda^2  |\chi_c|^2 \,\left[J_{\phi_{\eta c}} \bar J_{\phi_{\eta c}}+ J_{\theta_{\eta c}} \bar J_{\theta_{\eta c}}\right]_{(\zeta_c,\bar \zeta_c)}+ \frac{\lambda^2}{2}  (\chi_c)^2\, \left[J^2_{\phi_{\eta c}}+J^2_{\theta_{\eta c}}\right]_{(\zeta_c,\bar \zeta_c)}+\frac{\lambda^2}{2} (\bar \chi_c)^2\,  \left[\bar J^2_{\phi_{\eta c}}+\bar J^2_{\theta_{\eta c}}\right]_{(\zeta_c,\bar \zeta_c)}\nn
&= - \,\frac{\lambda^2}{4}  (x^2+u_c^2\,\tau^2) \,\Big[(\partial_X \phi_{\eta c})^2+\frac{(\partial_\mathcal{T} \phi_{\eta c})^2}{u_c^2}+(\partial_X \theta_{\eta c})^2+\frac{(\partial_\mathcal{T} \theta_{\eta c})^2}{u_c^2}\,\Big]\nn
&+\frac{\lambda^2}{8}  (-x^2+u_c^2\,\tau^2-2iu_cx\tau) \Big[(\partial_X \phi_{\eta c})^2-\frac{(\partial_\mathcal{T} \phi_{\eta c})^2}{u_c^2}+ \frac{2\,i\,(\partial_X \phi_{\eta c}) (\partial_\mathcal{T} \phi_{\eta c})}{u_c }+(\partial_X \theta_{\eta c})^2-\frac{(\partial_\mathcal{T} \theta_{\eta c})^2}{u_c^2}+ \frac{2\,i\,(\partial_X \theta_{\eta c}) (\partial_\mathcal{T} \theta_{\eta c})}{u_c}\Big]\nn
&+ \frac{\lambda^2}{8}(-x^2+u_c^2\,\tau^2+2iu_cx\tau)\Big[(\partial_X \phi_{\eta c})^2-\frac{(\partial_\mathcal{T} \phi_{\eta c})^2}{u_c^2}- \frac{2\,i\,(\partial_X \phi_{\eta c}) (\partial_\mathcal{T} \phi_{\eta c})}{u_c} +(\partial_X \theta_{\eta c})^2-\frac{(\partial_\mathcal{T} \theta_{\eta c})^2}{u_c^2}- \frac{2\,i\,(\partial_X \theta_{\eta c}) (\partial_\mathcal{T} \theta_{\eta c})}{u_c}  \Big]\nn
&= -\frac{\lambda^2}{2}\,\Big[x^2\,(\partial_X \phi_{\eta c})^2+\tau^2(\partial_\mathcal{T} \phi_{\eta c})^2+x^2\,(\partial_X \theta_{\eta c})^2+\tau^2(\partial_\mathcal{T} \theta_{\eta c})^2\Big].
\label{20}
\end{align} 
We note that the odd terms in $x$ and $\tau$ do not contribute to the RG equations. 
Next, we put $\lambda=1/\sqrt{2}$ in Eq. (\ref{20}) and make use of it in Eq. (\ref{19}) to get
\begin{align}
I^{d\alpha}= -\sum_\eta\frac{(\tilde t_F)^2\,u_c^2}{16\,\pi^2\, \alpha^4}\int dx\, dt\, dX\, d\mathcal{T}\,\frac{x^2\,(\partial_X \phi_{\eta c})^2+\tau^2(\partial_\mathcal{T} \phi_{\eta c})^2+x^2\,(\partial_X \theta_{\eta c})^2+\tau^2(\partial_\mathcal{T} \theta_{\eta c})^2}{(|\chi_+|/\alpha)^{ \nu_+} (|\chi_-|/\alpha)^{ \lambda_-}(|\chi_s|/\alpha)^{ \nu_s/2}}.
\end{align}
We again change to polar coordinates with $(x,u_c \tau)=(r\cos\theta',r\sin\theta')$ such that $I^{d\alpha}$ reads
\begin{align}
I^{d\alpha}&= - \frac{(\tilde t_F)^2\,u_c}{16\,\pi^2\, \alpha^4} \sum_\eta \int dX\, d\mathcal{T}\, \int\, r\,dr\,d\theta\,\frac{r^2\,\cos^2\theta'\left[(\partial_X \phi_{\eta c})^2+(\partial_X \theta_{\eta c})^2\right]+\frac{r^2\,\sin^2(\theta')}{u_c^2}\left[(\partial_\mathcal{T} \phi_{\eta  c})^2+(\partial_\mathcal{T} \theta_{\eta  c})^2\right]}{(|\chi_+|/\alpha)^{ \nu_+} (|\chi_-|/\alpha)^{ \lambda_-}(|\chi_s|/\alpha)^{ \nu_s/2}}\nn [1em]
&=-\frac{(\tilde t_F)^2\,u_c}{4} \Big(\frac{d\alpha}{\alpha}\Big)\sum_\eta \int \frac{dX\, d\mathcal{T}}{2\,\pi}\,\bigg(  \left[(\partial_X \phi_{\eta  c})^2+(\partial_X \theta_{\eta  c})^2 \right] h_x^{t_F} + \frac{\left[(\partial_\mathcal{T} \phi_{\eta c})^2+(\partial_\mathcal{T} \theta_{\eta c})^2 \right]\, h_t^{t_F}}{u_c^2}\bigg),
\label{25}
\end{align}
where $h_x^{t_F}$ and $h_\tau^{t_F}$ are defined by
\begin{align}
2\,\pi\,h_x^{t_F}=& \int_0^{2\,\pi} \frac{ \cos^2\theta' \,d\theta'  }{\left(\cos^2\theta'+\frac{u_+^2}{u_c^2} \sin^2\theta'\right)^{ \nu_+/2 } \left(\cos^2\theta'+\frac{u_-^2}{u_c^2} \sin^2\theta'\right)^{ \lambda_-/2}\left(\cos^2\theta'+\frac{u_s^2}{u_c^2} \sin^2\theta'\right)^{\nu_s/4}},\nn
2\,\pi\,h_\tau^{t_F}=&\int_0^{2\,\pi} \frac{\sin^2\theta' \,d\theta'}{\left(\cos^2\theta'+\frac{u_+^2}{u_c^2} \sin^2\theta'\right)^{ \nu_+/2} \left(\cos^2\theta'+\frac{u_-^2}{u_c^2} \sin^2\theta'\right)^{ \lambda_-/2}\left(\cos^2\theta'+\frac{u_s^2}{u_c^2} \sin^2\theta'\right)^{\nu_s/4}}.
\end{align}
From the equation of motion, we obtain $\partial_{\mathcal{T}} \theta_{\eta c}=-i \left[\frac{u_c}{K_c}\partial_X \phi_{\eta c}+ \tilde u\, \partial_X \phi_{\bar \eta c}\right]$, $\partial_{\mathcal T} \theta_{\eta s}=-i\,\frac{u_s}{K_s}\partial_X \phi_{\eta s}$ and $\partial_{\mathcal T} \phi_{\eta \nu}=-i\,u_{\nu}\,K_{\nu}\, \partial_X \theta_{\eta \nu}$. Therefore, Eq. (\ref{25}) reduces to 
\begin{align}
I^{d\alpha}=&-\frac{(\tilde t_F)^2\,u_c}{4} \Big(\frac{d\alpha}{\alpha}\Big) \int \frac{dX\, d\mathcal{T}}{2\,\pi}\,\bigg(  \left[(\partial_X \phi_{1  c})^2+(\partial_X \phi_{\bar 1  c})^2+(\partial_X \theta_{1  c})^2 + (\partial_X \theta_{\bar 1  c})^2\right] h_x^{t_F}-K_c^2\,[(\partial_X \theta_{1 c})^2+ (\partial_X \theta_{\bar 1 c})^2]\, h_\tau^{t_F} \nn 
&- \big[ \frac{(u_c^2/K_c^2+\tilde u^2)(\partial_X \phi_{ 1 c})^2+(u_c^2/K_c^2+\tilde u^2)(\partial_X \phi_{\bar 1 c})^2}{u_c^2} +\frac{4\, \tilde u\,(\partial_X \phi_{ 1 c})\,(\partial_X \phi_{\bar 1 c})}{K_c\,u_c}\big]\, h_\tau^{t_F}\bigg)\nn[1em]
&=-\frac{(\tilde t_F)^2\,u_c}{4} \Big(\frac{d\alpha}{\alpha}\Big) \int \frac{dX\, d\mathcal{T}}{2\,\pi}\,\Bigg(\left[h_x^{t_F}-\left(\frac{u_c^2}{K_c^2}+\tilde u^2\right)\, h_\tau^{t_F}\right]  (\partial_X \phi_{1  c})^2+\left[h_x^{t_F}-\left(\frac{u_c^2}{K_c^2}+\tilde u^2\right)\, h_\tau^{t_F}\right] (\partial_X \phi_{\bar 1  c})^2\nn
&+\left[h_x^{t_F}-K_c^2\,h_\tau^{t_F}\right]\,(\partial_X \theta_{1  c})^2 +\left[h_x^{t_F}-K_c^2\,h_\tau^{t_F}\right]\, (\partial_X \theta_{\bar 1  c})^2-\frac{4\, \tilde u\,(\partial_X \phi_{ 1 c})\,(\partial_X \phi_{\bar 1 c})}{K_c\,u_c}\, h_\tau^{t_F} \Bigg).
\label{27}
\end{align}
As a result, the RG equations corresponding to $\phi_{1c}$/$\theta_{1c}$ fields take the following form:
\begin{align}
&\frac{d}{dl} \left(\frac{u_c}{K_c}\right)= \frac{(\tilde t_F)^2 \,u_c\, \left[h_x^{t_F}-\left(\frac{u_c^2}{K_c^2}+\tilde u^2\right)\, h_\tau^{t_F}\right]}{4}, \,\,\frac{d}{dl}(u_c\,K_c) = \frac{(\tilde t_F)^2\,u_c  \left[h_x^{t_F}-K_c^2 h_\tau^{t_F}\right]}{4}, \frac{d\tilde u}{dl} =  -\frac{\tilde u}{2\,K_c}(\tilde t_F)^2\, h_\tau^{t_F},
\label{u2}\\
&\frac{du_c}{dl}=\frac{(\tilde t_F)^2}{8}\left(h_x^{t_F}-h_\tau^{t_F}\right)\,\left(u_c\,K_c+\frac{u_c}{K_c} \right)-\frac{(\tilde t_F)^2\,\tilde u^2\,K_c\,h_\tau^{t_F}}{8\,u_c}, \\ & \frac{dK_c}{dl}=\frac{(\tilde t_F)^2 }{8}\left(h_x^{t_F}+h_\tau^{t_F}\right)\,\left(1-K_c^2\right)+\frac{(\tilde t_F)^2\,\tilde u^2\,K_c^2\,h_\tau^{t_F}}{8\,u_c^2}.
\end{align}

For the spin-sector (corresponding to $\phi_{\eta s}$--$\theta_{\eta s}$), as follows from Eq. (\ref{16}), we obtain the form of $I_s$ as 
\begin{align}
I_s&= \frac{(\tilde t_F)^2\,u_c^2}{2\,\pi^2\, \alpha^4} \sum_\eta\int dx \,d\tau\, dX\, dT \frac{1}{2(|\chi_+|/\alpha)^{ 2\,\lambda^2\nu_+} (|\chi_-|/\alpha)^{ 2\,\lambda^2\lambda_-}(|\chi_s|/\alpha)^{\lambda^2 \nu_s}}\nn
&\hspace*{0.5cm}\times \bigg(-\lambda^2  |\chi_s|^2 \,\big[J_{\phi_{\eta s}} \bar J_{\phi_{\eta s}}+J_{\theta_{\eta s}} \bar J_{\theta_{\eta s}}\big]_{(\zeta_s,\bar \zeta_s)}+ \frac{\lambda^2\,(\chi_s)^2}{2}  \big[J^2_{\phi_{\eta s}}+J^2_{\theta_{\eta s}}\big]_{(\zeta_s,\bar \zeta_s)}+\frac{\lambda^2\,(\bar \chi_s)^2 }{2} \big[\bar J^2_{\phi_{\eta s}}+\bar J^2_{\theta_{\eta s}}\big]_{(\zeta_s,\bar \zeta_s)}
\bigg).
\end{align} 
We  repeat the same steps as before for Eqs. (\ref{20})-(\ref{27}) and after converting the integral to polar coordinates $(x,u_c \tau)=(r\cos\theta',r\sin\theta')$ and integrating $r$ from $\alpha$ to $\alpha+ d \alpha$, $I_s^{d\alpha}$ reads
\begin{align}
I_s^{d\alpha}&= -\frac{(\tilde t_F)^2 \,u_c\,\lambda^2}{2}\left( \frac{d\alpha}{\alpha}\right) \sum_\eta \int \frac{ dX d\mathcal T}{2\,\pi} \Big([(\partial_X \phi_{\eta s})^2+(\partial_X \theta_{\eta s})^2]\, h_x^{t_F} + \frac{[(\partial_T \phi_{\eta s})^2+(\partial_T \theta_{\eta s})^2]}{u_c^2} h_\tau^{t_F}\Big)\nn
&=-\frac{(\tilde t_F)^2 \,u_c\,\lambda^2}{2}\left( \frac{d\alpha}{\alpha}\right) \sum_\eta \int \frac{ dX d\mathcal T}{2\,\pi} \Big([h_x^{t_F}-\frac{u_s^2}{u_c^2\,K_s^2} h_\tau^{t_F}](\partial_X \phi_{\eta s})^2+[h_x^{t_F}-\frac{u_s^2\,K_s^2}{u_c^2}h_\tau^{t_F}](\partial_X \theta_{\eta s})^2\Big).
 \end{align}

Therefore, for $\lambda=1/\sqrt{2}$, we obtain
\begin{align}
&\frac{d}{dl} \left(\frac{u_s}{K_s}\right)= \frac{(\tilde t_F)^2 \,u_c\, [h_x^{t_F}-\frac{u_s^2}{u_c^2\,K_s^2} h_\tau^{t_F}]}{4}, \,\,\frac{d}{dl} \left(u_s\,K_s\right)= \frac{(\tilde t_F)^2\,u_c\, [h_x^{t_F}-\frac{u_s^2\,K_s^2}{u_c^2}h_\tau^{t_F}] }{4}, \nn
&\frac{du_s}{dl}=\frac{(\tilde t_F)^2}{8}\left(h_x^{t_F}-\frac{u_s^2}{u_c^2} \,h_\tau^{t_F}\right) \,\left(u_c\,K_s+ \frac{u_c}{K_s}\right) , \, \, \frac{dK_s}{dl}=\frac{(\tilde t_F)^2 }{8}\left(\frac{u_c}{u_s}\,h_x^{t_F}+\frac{u_s}{u_c} \,h_\tau^{t_F}\right)\,\left(1-K_s^2\right).
\end{align}

\subsubsection{Terms contributing to the interband crossed terms between the NWs : $(\partial_x \phi_{\eta c} \,\partial_x \phi_{\bar \eta c})$ and $(\partial_x \theta_{\eta c} \,\partial_x \theta_{\bar \eta c})$}
In this subsection, we calculate the terms generated under the RG flow and contributing to the interband cross term defined in Eq. (4). From the kinetic energy term calculations, we already get the contribution coming out from $\tilde\Delta_{sc}$ and $\tilde t_F$ and defined in Eqs. (\ref{u1}) and (\ref{u2}). Apart from these contributions, $\tilde t_F$ has direct terms as well as terms which contribute to the interband cross term. We collect this type of terms from Eq. (\ref{16}) and write the term similar to Eq. (\ref{19}) as
\begin{align}
I^{d\alpha}_{cr}&=\int dx \,dt\, dX\, d\mathcal{T} \frac{(\tilde t_F)^2\,u_c^2}{2\,\pi^2\, \alpha^4}\frac{1}{2(|\chi_+|/\alpha)^{ 2\,\lambda^2\nu_+} (|\chi_-|/\alpha)^{ 2\,\lambda^2\lambda_-}(|\chi_s|/\alpha)^{\lambda^2 \,\nu_s}}\nn
& \hspace*{0.75cm}\times \Big(- \lambda^2 |\chi_c|^2 [J_{\phi_{1c}} \bar J_{\phi_{\bar 1c}}+J_{\phi_{\bar 1c}} \bar J_{\phi_{1c}}-J_{\theta_{1c}} \bar J_{\theta_{\bar 1c}}-J_{\theta_{\bar 1c}} \bar J_{\theta_{ 1c}}]_{(\zeta_c,\bar \zeta_c)}+ \lambda^2(\chi_c)^2\, [J_{\phi_{1c}} J_{\phi_{\bar 1c}}]_{(\zeta_c,\bar \zeta_c)} \nn
&\hspace*{0.75cm}-\lambda^2 (\chi_c)^2\, [J_{\theta_{1c}} J_{\theta_{\bar 1c}}]_{(\zeta_c,\bar \zeta_c)}+\lambda^2 (\bar\chi_c)^2 \,[\bar J_{\phi_{1c}} \bar J_{\phi_{\bar 1c}}]_{(\zeta_c,\bar \zeta_c)} -\lambda^2 (\bar\chi_c)^2\, [\bar J_{\theta_{1c}} \bar J_{\theta_{\bar 1c}}]_{(\zeta_c,\bar \zeta_c)}\Big),
\end{align} 
To simplify the expression of $I^{d\alpha}_{cr}$, we calculate the explicit form of terms containing $J_{\phi_{\eta c}/\theta_{\eta c}}$ and $\bar J_{\phi_{\eta c}/\theta_{\eta c}}$ as
\begin{align}
&- \lambda^2 |\chi_c|^2 [J_{\phi_{1c}} \bar J_{\phi_{\bar 1c}}+J_{\phi_{\bar 1c}} \bar J_{\phi_{1c}}-J_{\theta_{1c}} \bar J_{\theta_{\bar 1c}}-J_{\theta_{\bar 1c}} \bar J_{\theta_{ 1c}}]_{(\zeta_c,\bar \zeta_c)}+ \lambda^2(\chi_c)^2\, [J_{\phi_{1c}} J_{\phi_{\bar 1c}}- J_{\theta_{1c}} J_{\theta_{\bar 1c}}]_{(\zeta_c,\bar \zeta_c)}\nn
&+\lambda^2 (\bar \chi_c)^2 \,[\bar J_{\phi_{1c}} \bar J_{\phi_{\bar 1c}}-\bar J_{\theta_{1c}} \bar J_{\theta_{\bar 1c}}]_{(\zeta_c,\bar \zeta_c)}
\nn
&= - \,\frac{\lambda^2}{2}  (x^2+u_c^2\,\tau^2) \,\Big[(\partial_X \phi_{1 c})(\partial_X \phi_{\bar 1 c})+ \frac{(\partial_\mathcal{T} \phi_{1 c})(\partial_\mathcal{T} \phi_{\bar 1 c})}{u_c^2}-(\partial_X \theta_{1 c})(\partial_X \theta_{\bar 1 c})-\frac{(\partial_\mathcal{T} \theta_{1 c})(\partial_\mathcal{T} \theta_{\bar 1 c})}{u_c^2}\,\Big]\nn
&\hspace*{0.4cm}+\frac{\lambda^2}{4}  (-x^2+u_c^2\,\tau^2-2\,i\,u_c\,x\,\tau) \Big[(\partial_X \phi_{1 c})(\partial_X \phi_{\bar 1 c})-\frac{(\partial_\mathcal{T} \phi_{1 c})(\partial_\mathcal{T} \phi_{\bar 1 c})}{u_c^2}+ \frac{i\,[(\partial_X \phi_{1 c}) (\partial_\mathcal{T} \phi_{\bar 1 c})+(\partial_\mathcal{T} \phi_{1 c}) (\partial_X \phi_{\bar 1 c})]}{u_c }\nn
&\hspace*{0.4cm}-(\partial_X \theta_{1 c})(\partial_X \theta_{\bar 1 c})+\frac{(\partial_\mathcal{T} \theta_{1 c})(\partial_\mathcal{T} \theta_{\bar 1 c})}{u_c^2}- \frac{i\,[(\partial_X \theta_{1 c}) (\partial_\mathcal{T} \theta_{\bar 1 c})+(\partial_\mathcal{T} \theta_{1 c}) (\partial_X \theta_{\bar 1 c})]}{u_c}\Big]\nn
&\hspace*{0.4cm}+\frac{\lambda^2}{4}  (-x^2+u_c^2\,\tau^2+2\,i\,u_c\,x\,\tau) \Big[(\partial_X \phi_{1 c})(\partial_X \phi_{\bar 1 c})-\frac{(\partial_\mathcal{T} \phi_{1 c})(\partial_\mathcal{T} \phi_{\bar 1 c})}{u_c^2}- \frac{i\,[(\partial_X \phi_{1 c}) (\partial_\mathcal{T} \phi_{\bar 1 c})+(\partial_\mathcal{T} \phi_{1 c}) (\partial_X \phi_{\bar 1 c})]}{u_c }\nn
&\hspace*{0.4cm}-(\partial_X \theta_{1 c})(\partial_X \theta_{\bar 1 c})+\frac{(\partial_\mathcal{T} \theta_{1 c})(\partial_\mathcal{T} \theta_{\bar 1 c})}{u_c^2}+ \frac{i\,[(\partial_X \theta_{1 c}) (\partial_\mathcal{T} \theta_{\bar 1 c})+(\partial_\mathcal{T} \theta_{1 c}) (\partial_X \theta_{\bar 1 c})]}{u_c}\Big]\nn
&\approx-\lambda^2\,x^2\,(\partial_X \phi_{1 c})(\partial_X \phi_{\bar 1 c})+\lambda^2\tau^2\,(\partial_\mathcal{T} \theta_{1 c})(\partial_\mathcal{T} \theta_{\bar 1 c})\nn
&=-\lambda^2\,\left[x^2+\frac{u_c^2\,\tau^2}{K_c^2}+\tilde u^2\,\tau^2\right]\,(\partial_X \phi_{1 c})(\partial_X \phi_{\bar 1 c})- \frac{ \lambda^2\tau^2\,\tilde u\,u_c}{K_c}\left[(\partial_X \phi_{1 c})^2+(\partial_X \phi_{\bar 1 c})^2\right].
\end{align} 
Here, we consider the term proportional to $(\partial_X\phi_{1 c})(\partial_X\phi_{\bar 1 c})$ and $(\partial_{\mathcal T}\theta_{1 c})(\partial_{\mathcal T}\theta_{\bar 1 c})$ and utilized the definition from the equation of motion: $\partial_{\mathcal T} \theta_{\eta c}=-i [\frac{u_c}{K_c}\partial_X \phi_{\eta c}+ \tilde u\, \partial_X \phi_{\bar \eta c}]$. Therefore, for $\lambda=1/\sqrt{2}$, we obtain
\begin{align}
I^{d\alpha}_{cr}= -\frac{(\tilde t_F)^2\,u_c^2}{8\,\pi^2\, \alpha^4}\int dx\, d\tau\, dX\, d{\mathcal T}\,\frac{ [x^2+\frac{u_c^2\,\tau^2}{K_c^2}+\tilde u^2\,\tau^2](\partial_X \phi_{1 c})(\partial_X \phi_{\bar 1 c})+ \frac{ \tau^2\,\tilde u\,u_c}{K_c}[(\partial_X \phi_{1 c})^2+(\partial_X \phi_{\bar 1 c})^2]}{(|\chi_+|/\alpha)^{ \nu_+} (|\chi_-|/\alpha)^{ \lambda_-}(|\chi_s|/\alpha)^{ \nu_s/2}}.
\end{align}
Further, converting the integral to polar coordinates and integrating from $\alpha$ to $\alpha+d \alpha$, we find
\begin{align}
I^{d\alpha}_{cr}&= -\frac{(\tilde t_F)^2\,u_c}{8\,\pi^2\, \alpha^4}\int r\,dr\, d\theta'\, dX\, d{\mathcal T}\,\frac{ 1}{(|\chi_+|/\alpha)^{ \nu_+} (|\chi_-|/\alpha)^{ \lambda_-}(|\chi_s|/\alpha)^{ \nu_s/2}}\\
&\times \bigg(\left[r^2\,\cos^2(\theta')+\frac{r^2\,\sin^2(\theta')}{K_c^2}+\frac{\tilde u^2\,r^2\,\sin^2(\theta')}{u_c^2}\right](\partial_X \phi_{1 c})(\partial_X \phi_{\bar 1 c})+ \frac{ r^2\,\sin^2(\theta')\,\tilde u}{u_c\,K_c}\left[(\partial_X \phi_{1 c})^2+(\partial_X \phi_{\bar 1 c})^2\right]\bigg)\nn[1em]
&=-\frac{(\tilde t_F)^2\,u_c}{2}\left(\frac{d\alpha}{\alpha}\right)\int  \frac{dX\, d{\mathcal T}}{2\pi}\,\bigg( \left[h_x^{t_F}+ h_\tau^{t_F}\left(\frac{1}{K_c^2}+\frac{\tilde u^2}{u_c^2}\right)\right] (\partial_X \phi_{1 c})(\partial_X \phi_{\bar 1 c})+ \frac{ h_\tau^{t_F}\tilde u}{u_c\,K_c}\left[(\partial_X \phi_{1 c})^2+(\partial_X \phi_{\bar 1 c})^2\right]\bigg). \nonumber
\end{align}
\begin{figure}[t]
\begin{center}
\includegraphics[width=0.45\columnwidth]{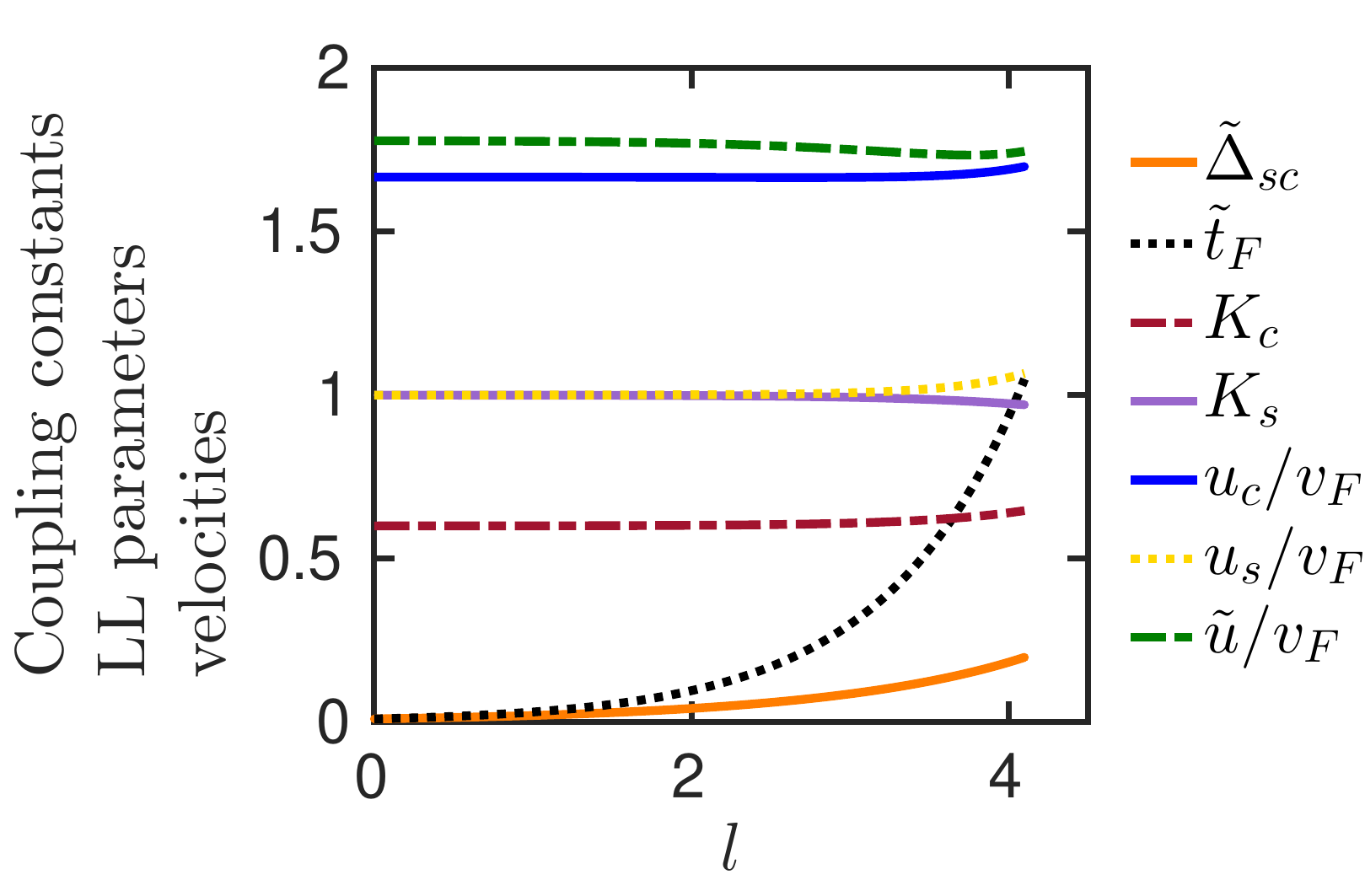}
\caption{The RG flow of different coupling constants, LL parameter, and velocities as function of the flow parameter $l$, with the initial conditions: $\tilde \Delta_{sc}(0)= \tilde t_F(0)=0.01, K_c(0)=0.6, K_s(0)=1,  u_{\nu}(0)/v_F=1/K_{\nu}(0)$, and $\tilde u(0)/v_F= 1/K_c^2(0)-1$, where $v_F$ is the Fermi velocity. The dimensionless Floquet Zeeman term (black dotted) reaches the strong coupling limit much faster compared to the superconducting pairing gap $\tilde \Delta_{sc}$ (orange solid), which results in the topological phase for the given initial parameter values at $l=0$. } 
\label{RG_flow}
\end{center}
\end{figure}

This generates the following contribution to the RG equations:
\begin{align}
&\frac{d \tilde{u}}{dl} = \frac{(\tilde{t}_F)^2 u_c}{4}\left[h_x^{t_F}+ h_\tau^{t_F}\left(\frac{1}{K_c^2}+\frac{\tilde u^2}{u_c^2}\right)\right],\\
&\frac{d }{dl}\left( \frac{u_c}{K_c}\right)= \frac{(\tilde{t}_F)^2\tilde u\,h_\tau^{t_F}}{2\,K_c},\, \,\frac{d }{dl}\left( u_c\,K_c\right)=0,\nn &
\frac{d u_c}{dl}=\frac{(\tilde{t}_F)^2 \tilde u \,h_\tau^{t_F}}{4}, \, \,\frac{d K_c}{dl}= -\frac{(\tilde{t}_F)^2\, \tilde u\,K_c\,h_\tau^{t_F}}{4\,u_c}.
\end{align}

To conclude, we collect all the terms from the Hamiltonians $H'_{sc}$ and $H'_Z$, which are contributing to the dimensionless coupling constants, LL parameters, and velocities and obtain the final form of the RG equations:

\begin{align}
&\frac{d}{dl}\tilde \Delta_{sc}= \;\left[2-\frac{\lambda_+ +\lambda_-+ K_s}{2}\right]\,\tilde \Delta_{sc},\nn 
&\frac{d\tilde t_F}{dl}= \;\left[2-\frac{ 2\,\nu_++ 2\,\lambda_- + \nu_s }{4}\right]\, \tilde t_F,\nn 
&\frac{d K_c}{dl} = \;\left[h_x^\Delta + \left( 1+\frac{K_c^2\,\tilde u^2}{u_c^2}\right)h_t^\Delta\right](\tilde{\Delta}_{sc})^2+\frac{\left(1 - K_c^2 \right)h_x^{t_F} + \left[\left(1 - \frac{\tilde{u}}{u_c}K_c \right)^2 - K_c^2  \right]h_t^{t_F}}{8}\,(\tilde{t}_F)^2,
\nn 
&\frac{d K_s}{dl} =\;-\frac{\left(\frac{u_c}{u_s}h_x^\Delta + \frac{u_s}{u_c}h_t^\Delta\right)}{2}\,K_s^2\, [(\tilde{\Delta}_{sc})^2+(\tilde{\Delta}^{ext})^2] + \frac{\left(\frac{u_c}{u_s}h_x^{t_F} + \frac{u_s}{u_c} h_t^{t_F} \right) \left(1 - K_s^2 \right)  }{8}\,(\tilde{t}_F)^2, \nn 
&\frac{du_c}{dl} =\; \frac{\left[h_x^\Delta - \left( 1+\frac{K_c^2\,\tilde u^2}{u_c^2}\right)h_t^\Delta\right]}{2\,K_c}\,u_c\,[(\tilde{\Delta}_{sc})^2+(\tilde{\Delta}^{ext})^2]+\frac{\left(1+K_c^2 \right)h_x^{t_F} -\left[\left(1 - \frac{\tilde{u}}{u_c}K_c \right)^2 + K_c^2 \right]h_t^{t_F}}{8\,K_c}\,u_c\, (\tilde{t}_F)^2, \nn 
&\frac{du_s}{dl} =\; \frac{\left(h_x^\Delta - \frac{u_s^2}{u_c^2}h_t^\Delta\right)}{2}\,u_c\, K_s\, [(\tilde{\Delta}_{sc})^2+(\tilde{\Delta}^{ext})^2]+\frac{\left(h_x^{t_F} - \frac{u_s^2}{u_c^2} h_t^{t_F} \right)\left(1 + K_s^2 \right) }{8\,K_s}\,u_c\,(\tilde{t}_F)^2, \nn 
&\frac{d\tilde{u}}{dl} =\;- \frac{2\,\tilde u\, h_t^{\Delta}}{K_c} [(\tilde{\Delta}_{sc})^2+(\tilde{\Delta}^{ext})^2]+\frac{\left[h_x^{t_F}+\left(\frac{1}{K_c} - \frac{\tilde{u}}{u_c} \right)^2 h_t^{t_F} \right]}{4}\,u_c\,(\tilde{t}_F)^2, 
\label{finalRG1}
\end{align}
where we define
\begin{align}
h_x^\Delta=& \,\frac{1}{2\,\pi}\int_0^{2\,\pi} \frac{ \cos^2\theta' \,d\theta'  }{\left(\cos^2\theta'+\frac{u_+^2}{u_c^2} \sin^2\theta'\right)^{ \lambda_+/2} \left(\cos^2\theta'+\frac{u_-^2}{u_c^2} \sin^2(\theta')\right)^{ \lambda_-/2}\left(\cos^2\theta'+\frac{u_s^2}{u_c^2} \sin^2\theta'\right)^{ K_s/2}},\nn 
h_t^\Delta=&\,\frac{1}{2\,\pi}\int_0^{2\,\pi} \frac{\sin^2\theta' \,d\theta'  }{\left(\cos^2\theta'+\frac{u_+^2}{u_c^2} \sin^2\theta'\right)^{ \lambda_+/2} \left(\cos^2\theta'+\frac{u_-^2}{u_c^2} \sin^2\theta'\right)^{\lambda_-/2}\left(\cos^2\theta'+\frac{u_s^2}{u_c^2} \sin^2\theta'\right)^{ K_s/2}},\nn 
h_x^{t_F}=&\, \frac{1}{2\,\pi}\int_0^{2\,\pi} \frac{ \cos^2\theta' \,d\theta'  }{\left(\cos^2\theta'+\frac{u_+^2}{u_c^2} \sin^2\theta'\right)^{ \nu_+/2 } \left(\cos^2\theta'+\frac{u_-^2}{u_c^2} \sin^2(\theta')\right)^{ \lambda_-/2}\left(\cos^2\theta'+\frac{u_s^2}{u_c^2} \sin^2\theta'\right)^{\nu_s/4}},\nn 
h_t^{t_F}=& \,\frac{1}{2\,\pi}\int_0^{2\,\pi} \frac{\sin^2\theta' \,d\theta'}{\left(\cos^2\theta'+\frac{u_+^2}{u_c^2} \sin^2\theta'\right)^{ \nu_+/2} \left(\cos^2\theta'+\frac{u_-^2}{u_c^2} \sin^2\theta'\right)^{ \lambda_-/2}\left(\cos^2\theta'+\frac{u_s^2}{u_c^2} \sin^2\theta'\right)^{\nu_s/4}},\nn
\lambda_{\pm}=&\,\frac{1}{2K_c} \frac{u_c \pm \tilde{u}\,K_c}{u_\pm}, \;\;\;\nu_+= \frac{K_c\, u_c}{2\,u_+}, \;\;\; \nu_s= \left(K_s+ \frac{1}{ K_s}\right)\;\;\; u_{\pm}^2 =\,u_c^2 \pm u_c \, \tilde{u}\, K_c.
\label{finalRG2}
\end{align}

The  coupled RG equations Eq. \ref{finalRG1}  are solved numerically and the coupling constants, LL parameters, and velocities are plotted as a function of $l$ in Fig. \ref{RG_flow}. The charge-spin velocities and LL parameter stay close to their initial values at $l=0$, which again verifies the validity of the assumption used for the analytical calculations performed in the main text of the paper.

\end{document}